\documentclass[aps,pre,groupedaddress,11pt,tightenlines]{revtex4-1}
\usepackage{graphicx}
\usepackage{amssymb}
\usepackage{amsmath}
\usepackage{epstopdf}
\usepackage[caption=false]{subfig}

\begin{document}

\title{A Three Dimensional Gravitational Billiard in a Cone}

\author{Cameron K. Langer}
\email{cklanger@ucdavis.edu}
\affiliation{Department of Physics and Astronomy, Texas Christian University}
\author{Bruce N. Miller}
\email{b.miller@tcu.edu}
\affiliation{Department of Physics and Astronomy, Texas Christian University}

\begin{abstract}
Billiard systems offer a simple setting to study regular and chaotic dynamics. Gravitational billiards are generalizations of these classical billiards which are amenable to both analytical and experimental investigations. Most previous work on gravitational billiards has been concerned with two dimensional boundaries. In particular the case of linear boundaries, also known as the wedge billiard, has been widely studied. In this work, we introduce a three dimensional version of the wedge; that is, we study the nonlinear dynamics of a billiard in a constant gravitational field colliding elastically with a linear cone of half angle $\theta$. We derive a two-dimensional Poincar\'{e} map with two parameters, the half angle of the cone and $\ell$, the $z$-component of the billiard's angular momentum. Although this map is sufficient to determine the future motion of the billiard, the three-dimensional nature of the physical trajectory means that a periodic orbit of the mapping does not always correspond to a periodic trajectory in coordinate space. We demonstrate several integrable cases of the parameter values, and analytically compute the system's fixed point, analyzing the stability of this orbit as a function of the parameters as well as its relation to the physical trajectory of the billiard. Next, we explore the phase space of the system numerically. We find that for small values of $\ell$ the conic billiard exhibits behavior characteristic of two-degree-of-freedom Hamiltonian systems with a discontinuity, and the dynamics is qualitatively similar to that of the wedge billiard, although the correspondence is not exact. As we increase $\ell$ the dynamics becomes on the whole less chaotic, and the correspondence with the wedge billiard is lost.
\end{abstract}

\maketitle

\section{Introduction\label{sec:1}}
Billiard systems are a class of simple, analytically tractable models exhibiting the fundamental aspects of nonlinear dynamics. These systems, first introduced by Birkhoff~\cite{birkhoff1927}, consist of a point mass (the ``billiard'' or ``particle'') in a two-dimensional (convex) region with a piecewise smooth boundary, where the motion between collisions is inertial and collisions with the boundary are specular and elastic. The natural way to study these systems is using a Poincar\'{e} surface of section taken at encounters with the boundary; this reduces the dynamics to a discrete mapping. For different boundaries the motion can be regular, chaotic, or mixed with KAM islands \cite{berry1981}. For example, the stadium billiard has been proven by Bunimovich~\cite{bunimovich1974,bunimovich1979} to be ergodic while the elliptic billiard, first introduced by Berry~\cite{berry1981}, is integrable. In addition quantum versions of these systems have also been studied by Waalkens et al. \cite{waalkens1996} among others. While these classical billiards are optimal for analytical study, more experimentally approachable models accounting for the Earth's gravitational field, called \emph{gravitational billiards}, have also been widely studied \cite{miller86,richter1990,wojtkowski1990,whelan1990,korsch1991,1szeredi1993,2szeredi1993,miller1999,miller2013,dacosta2015}. In \cite{miller86}, the \emph{wedge billiard}, consisting of a particle falling between two symmetric linear boundaries of angle $2\theta$, was shown by Lehtihet and Miller to exhibit the full range of possible behavior in Hamiltonian systems with two degrees of freedom. Namely, for $\theta<45^\circ$ the phase space consists of a mixed phase space with regular and chaotic regions, for $\theta=45^\circ$ the system is integrable, and for $\theta>45^\circ$ the system is ergodic. Subsequent work on the wedge billiard includes that of Richter et al.~\cite{richter1990}, where the oscillations in the relative amount of chaotic versus regular parts of the phase space (the so-called ``breathing chaos'') are discussed in terms of the symmetry lines of the system, as well as Wojtkowski \cite{wojtkowski1990} who rigorously established the ergodicity of the wedge for $\theta>45^\circ$. The one-dimensional motion of two particles in a constant gravitational field, a system which is isomorphic to the wedge, was studied by Whelan et al.~\cite{whelan1990} in terms of the stability of fixed point orbits. Korsch and Lang \cite{korsch1991} demonstrated the integrability of the motion of a gravitational billiard in a parabola, and the corresponding hyperbolic system was investigated by Ferguson et al. in~\cite{miller1999}, where the wedge and parabolic behavior was shown to arise in two limiting cases of the parameter values. The quantum version of the wedge was discussed by Szeredi and Goodings~\cite{1szeredi1993,2szeredi1993} in terms of Gutzwiller's periodic orbit theory. More recently, the circular, elliptic, and oval gravitational billiards were studied by da Costa et al.~\cite{dacosta2015}, who showed that the energy of these systems plays a key role in separating regular and apparently ergodic behavior. For the case of elastic collisions, the possible dynamical behavior in two-dimensional Hamiltonian billiards is well established. \\\indent
Classical billiard systems have also been studied in three (and higher) dimensions. The integrable class of ellipsoidal and related billiards were introduced by Richter et al.~\cite{richter1995} via action integrals, while the semiclassical and quantum versions were studied by Waalkens et al.~\cite{waalkens1999}. However, less is known about three dimensional \emph{gravitational} billiards, as these are nonintegrable in general.\\\indent
One way to study billiards (both classical and quantum) experimentally is by bouncing ultracold atoms off of beams of light. These so-called ``optical billiards'' were introduced by Raizen et al.~\cite{raizen1999} and provide a testing ground for undriven gravitational billiards. Of course, in any macroscopic billiard energy is no longer conserved, so in order to observe nontrivial experimental behavior the \emph{driven} versions of these two-dimensional gravitational billiards must be considered. Recently, the driven wedge, parabolic, and hyperbolic gravitational billiards were studied experimentally in \cite{feldt2005}. In their experiment, Feldt and Olafsen used a steel ball moving within an aluminum wedge, parabolic, or hyperbolic boundary, bounded from above by another aluminum boundary. The boundary was driven sinusoidally in the horizontal direction to counteract energy loss due to collisions. The results of this experiment showed that the driven parabolic system was regular, the wedge chaotic, and the hyperbolic mixed, sharing (as in the static case) characteristics of the wedge and parabolic systems at different parameter values.\\\indent
One-dimensional driven gravitational systems have been thoroughly investigated~\cite{holmes1982,luck1993,vogel2011,okninski2009,okninski2009-2,langer2015}. The seminal example is the so-called \emph{gravitational bouncer}, consisting of a particle impacting a periodically driven wall in the presence of a constant gravitational field. The gravitational bouncer was introduced as a variant of the well-known Fermi-Ulam model~\cite{fermi1949,ulam1961}, and has been studied for several types of driving motions including sinusoidal~\cite{holmes1982,luck1993,vogel2011} and piecewise linear~\cite{okninski2009,okninski2009-2,langer2015}. To model the Feldt experiments, two-dimensional driven gravitational billiards were studied numerically in \cite{gorski2006,miller2013}; in the former theoretical study rotational effects were ignored, while in the latter rotational effects were included in the model, making analytical computations e.g., periodic orbits difficult. However, in the work of Hartl et al.~\cite{miller2013}, the numerical simulations supported the experimental results of~\cite{feldt2005}, except in secondary quantities derived from the data, such as the tangential velocity of the particle after collisions.\\\indent
 One possible issue with experiments conducted on the driven wedge is the two-dimensional nature of the idealized system. In the real system, the billiard is not a point particle and thus has rotational properties which are affected at each collision. In the experiments of Feldt et al., additional boundaries were used to ensure the motion of the billiard was contained in the plane; however, any collision with these boundaries would likely play a nontrivial role in the dynamics. One way of eliminating this problem would be to get rid of the constraint that the motion be contained in a plane. If, instead of a particle in a wedge, we considered a particle in a \emph{cone}, then there would be no need for additional boundaries. In fact, such a system would be ideal for studying the effects of rotation on billiard systems, as the cone could either be driven in the conventional sense (i.e., the entire cone oscillating in a fixed direction) or the cone could spin in a sinusoidal fashion. As the equations determining the time of the next collision are in a sense unaffected by this ``rotational'' driving, such a system is analytically tractable, while also being experimentally realizable.\\\indent 
In this paper, we take the first step to understanding the conic billiard system. Namely, we consider the motion of a uniformly accelerated particle in $\mathbb{R}^3$, colliding elastically with a linear cone of half-angle $\theta$. Along with the energy, the $z$-component of the particle's angular momentum is conserved. Using these two integrals of the motion together with the fact that the dynamics depends only on the difference in the azimuthal angle $\phi$ between collisions, the conic billiard is reduced to a two-dimensional Poincar\'{e} surface of section with two parameters. However, in order to determine the physical trajectory of the billiard uniquely the corresponding change in azimuthal angle must be accounted for as well.\\\indent
This paper is organized as follows. In section~\ref{sec:2}, we introduce the model and derive the discrete mapping characterizing it. In section~\ref{sec:3} we demonstrate some simple properties of the mapping, including integrable limits and the existence of periodic orbits. We analyze the linear stability of these periodic orbits in terms of the parameter values. In section~\ref{sec:4} we present some numerical results, and in section~\ref{sec:5} we discuss the results obtained, as well as possible extensions of our model.

\section{The model and the mapping\label{sec:2}}
As the motion of the particle is unaffected by its mass, we are free to set the mass of the particle equal to unity, without loss of generality. We orient our Cartesian coordinate system so that the axis of the cone is the $z$-axis, and $\theta$ is the usual spherical polar angle~\footnote{We use $\theta$ as both the polar angle \emph{and} the half-angle of the cone, because at a collision these angles must be equal; since we use a discrete map between collisions, the meaning of $\theta$ should be clear from context.}. The gravitational field is $\mathbf{g}=-g\mathbf{\hat{e}}_z$, where $\mathbf{\hat{e}}_z$ is a unit vector in the $z$-direction. We shall find that the Poincar\'{e} map is most easily obtained in spherical polar coordinates $(r,\theta,\phi)$. In this coordinate system, the Hamiltonian of the billiard between collisions is \begin{equation} H=\frac{1}{2}\left[p_r^2+\frac{1}{r^2}\left(p_\theta^2+\frac{p_\phi^2}{\sin^2\theta}\right)\right]+gr\cos\theta. \end{equation} This governs the motion of the particle between collisions, which we assume to be elastic. There are two independent integrals of the motion preserved by both the parabolic motion \emph{and} collisions between the particle and the cone: \begin{equation} H=E,\qquad\qquad\mathrm{and}\qquad\qquad p_\phi=\ell_z,\end{equation} which we identify as the energy and $z$-component of angular momentum, respectively. The presence of two conserved quantities reduces the dimension of the phase space of the system from six to four. Taking the natural Poincar\'{e} surface-of-section at the moment of collision (here $\rho=\sqrt{x^2+y^2}$ is the cylindrical polar coordinate), \begin{equation} z=\rho\cot\theta=r\cos\theta\end{equation} further reduces the dynamics to a three-dimensional map. In fact, as we shall show below, the azimuthal angle $\phi$ plays no role in the dynamics, and therefore the arbitrary initial condition $\phi=\phi_0$ is sufficient to describe all possible behavior of the system. Thus, a two-dimensional map characterizes the dynamics. By a suitable transformation of the coordinates and time, the constants $E$ and $g$ can be re-scaled arbitrarily; for convenience, we choose units such that $E=g=\frac{1}{2}$, so that the three constants $E,g,\ell_z$ are consolidated into the single (dimensionless) parameter (where the factor of $\frac{1}{\sqrt{2}}$ is chosen for aesthetic purposes) \begin{equation} \ell'\equiv\frac{g\ell_z}{\sqrt{2}E^{3/2}}.\end{equation} With this choice of units, energy and $z$-component of angular momentum conservation are expressed by (where we have replaced the momenta $p_r,p_\theta$ with the corresponding velocities $v_r,v_\theta$)  \begin{equation} 1=v_r^2+v_\theta^2+\frac{\ell'^2}{r^2\sin^2\theta}+r\cos\theta, \qquad \ell'=r v_\phi\sin\theta.\end{equation} The cone angle $\theta$ is restricted to $0^\circ\leqslant \theta\leqslant 90^\circ$. It is not difficult to show using energy conservation that $\ell'$ satisfies $\vert\ell'\vert\leqslant\tan\theta$. In fact, a more careful analysis shows that this bound can be lowered to $\vert\ell'\vert\leqslant\frac{2\tan\theta}{3\sqrt{3}}$; calling this upper bound $\ell_\mathrm{max}$, we obtain a parameter which ranges from zero to one by defining $\ell\equiv\frac{\ell'}{\ell_{\mathrm{max}}}$. This defines the system and its parameters.\\\indent
We have shown that a two dimensional map is sufficient to characterize the dynamics of the conic billiard. However, we not yet shown which variables are most suitable. It is clear that this choice cannot be made haphazardly, as most pairs of conjugate variables are insufficient to fully determine the particle's subsequent trajectory. As we shall see below, it turns out that the radial coordinate and the radial component of the velocity, $(r,v_r)$ provide enough information to determine the subsequent motion unambiguously (for a given choice of initial $\phi$). Thus, our aim is to compute the Poincar\'{e} map \begin{equation} \mathcal{P}:\begin{pmatrix} r_n \\ v_{n_r} \end{pmatrix}\rightarrow\begin{pmatrix} r_{n+1} \\ v_{{n+1}_r} \end{pmatrix},\end{equation} where the suffix $n$ denotes the value of a quantity just \emph{after} the $n$th collision. In order to compute this map, we require two ingredients: $(1)$ the time interval between collisions, and $(2)$ the law relating the pre-collision and post-collision components of the particle's velocity. To compute $(1)$, we need only basic kinematics; in Cartesian coordinates, the time of the $(n+1)$st collision is defined by the implicit relations \begin{equation} v_{n_x}(t_{n+1}-t_n)+x_n=x_{n+1},\qquad v_{n_y}(t_{n+1}-t_n)+y_n=y_{n+1},\end{equation} \begin{equation} -\frac{1}{4}(t_{n+1}-t_n)^2+v_{n_z}(t_{n+1}-t_n)+\sqrt{x_n^2+y_n^2}\cot\theta=\sqrt{x_{n+1}^2+y_{n+1}^2}\cot\theta
,\end{equation} where the factor $\frac{1}{4}$ is due to the fact that we set $g=\frac{1}{2}$, and we have used the relation $z=\rho\cot\theta$ defining the surface of the cone. These equations are equivalent to the cubic equation (where $\tau_{n+1}\equiv t_{n+1}-t_n$) \begin{multline} \tau_{n+1}^3-8v_{n_z}\tau_{n+1}^2+16\left[v_{n_z}^2-(v_{n_x}^2+v_{n_y}^2)\cot^2\theta-\frac{\sqrt{x_n^2+y_n^2}}{2}\cot\theta\right]\tau_{n+1} \\+32\left[v_{n_z}\sqrt{x_n^2+y_n^2}\cot\theta-\left(x_nv_{n_x}+y_nv_{n_y}\right)\cot^2\theta\right]=0,\end{multline} which can be expressed as \begin{multline}\label{eq:time}\tau_{n+1}^3+8(v_{n_\theta}\sin\theta-v_{n_r}\cos\theta)\tau_{n+1}^2\\+16\left\{v_{n_\theta}^2-\left[(v_{n_\theta}^2+v_{n_\phi}^2)\cot\theta+2v_{n_r}v_{n_\theta}\right]\cot\theta-\frac{r_n\cos\theta}{2}\right\}\tau_{n+1} \\-32r_n\,v_{n_\theta}\cot\theta=0\end{multline} in spherical coordinates. The smallest positive root of this cubic equation is the time of the $(n+1)$st collision. The roots of this cubic equation are long and complicated, and do not yield any physical insight since a priori there is no easy way to unambiguously determine the smallest positive of the three (real) roots. Thus, they are not included here.\\\indent
To analyze the effects of a collision, spherical coordinates are natural for this system. This is because the equations relating pre-collision and post-collision components of the billiard's velocity become especially simple in these coordinates, since $\mathbf{\hat{e}}_\theta$ is orthogonal to the surface of the cone, and $\mathbf{\hat{e}}_r,\mathbf{\hat{e}}_\phi$ are parallel to it (where $\mathbf{\hat{e}}_r,\mathbf{\hat{e}}_\theta,\mathbf{\hat{e}}_\phi$ are unit vectors in the direction of $r,\theta$ and $\phi$, respectively) at the moment of collision. If $v_{{n+1}_i}^-$ denotes the $i$th component of the particle's velocity just \emph{before} the $(n+1)$st collision, and $v_{{n+1}_i}^+$ the corresponding component just \emph{after} the $(n+1)$st collision, then the vector equation (here $\mathbf{\hat{n}}=\mathbf{\hat{e}}_\theta$ is a unit normal to the surface of the cone, and $\mathbf{v}^{+/-}_{n+1}$ is the velocity of the particle just after/before the $(n+1)$st collision) \begin{equation} \mathbf{v}^+_{n+1}=\mathbf{v}^-_{n+1}-2(\mathbf{v}^-_{n+1}{\cdot}\mathbf{\hat{n}})\mathbf{\hat{n}} \end{equation} is equivalent to the three component equations \begin{equation} v_{{n+1}_r}^+=v_{{n+1}_r}^-,\qquad v_{{n+1}_\phi}^+=v_{{n+1}_\phi}^-,\qquad v_{{n+1}_\theta}^+=-v_{{n+1}_\theta}^-. \end{equation} Now we demonstrate that knowledge of $r$ and $v_r$ at the $n$th collision is sufficient to describe the subsequent motion of the particle. Suppose these values, $(r_n,v_{n_r})$ at the $n$th collision are known. Then we immediately know the value of $\theta_n=\theta$, since at each collision this is just the half angle of the cone; additionally, we know the $\phi$-component of the particle's velocity, from the conservation of the $z$-component of angular momentum: \begin{equation} v_{n_\phi}=\frac{\ell}{r_n\sin\theta}.\end{equation} Using this in the energy conservation expression permits us to solve for $v_{n_\theta}^2$ in terms of $r_n$ and $v_{n_r}$: \begin{equation}\label{eq:energy} v_{n_\theta}^2=1-r_n\cos\theta-v_{n_r}^2-\frac{\ell^2}{r_n^2\sin^2\theta}.\end{equation} Although it seems we may only know $v_{n_\theta}$ up to a sign, in fact we are safe to choose the \emph{negative} root, since $v_{\theta}$ is the component of the velocity orthogonal to the surface of the cone in the direction of $\mathbf{\hat{e}}_\theta$, which, due to the convention $\mathbf{\hat{e}}_\theta\equiv\mathbf{\hat{e}}_\phi\times\mathbf{\hat{e}}_r$, cannot be \emph{positive} after a collision. Moreover, as we shall see below, the azimuthal angle $\phi_n$ plays no role in the dynamics; thus, it can be fixed arbitrarily at $t=0$ and subsequently eliminated from the mapping, as only the \emph{difference} $\phi_{n+1}-\phi_n$ appears in the mapping equations (and this quantity can be computed purely in terms of $r_n,v_{n_r}$ and the time of the next collision). Thus, to complete the mapping we need only compute the time interval from~\eqref{eq:time}, the new radial distance $r_{n+1}=\sqrt{x_{n+1}^2+y_{n+1}^2}\csc\theta$ and the radial velocity component $v_{{n+1}_r}$ from (see Appendix) \begin{multline}\label{eq:vel} v_{{n+1}_r}=v_{n_r}(\sin^2\theta\cos\varphi_{n+1}+\cos^2\theta)+v_{n_\theta}(\cos\varphi_{n+1}-1)\sin\theta\cos\theta\\+v_{n_\phi}\sin\varphi_{n+1}\sin\theta-\frac{1}{2}\tau_{n+1}\cos\theta,\end{multline} where $\varphi_{n+1}\equiv\phi_{n+1}-\phi_n$ is the difference in azimuthal angle between collisions, given by \begin{equation}\label{eq:phi} \varphi_{n+1}=\arctan\left[\frac{v_{n_\phi}\tau_{n+1}}{r_n\sin\theta+(v_{n_r}\sin\theta+v_{n_\theta}\cos\theta)\tau_{n+1}}\right],\end{equation} with care being taken to choose the correct quadrant in the $xy$-plane. In terms of these quantities the mapping for $r_{n+1}$ is \begin{equation} \label{eq:map} r_{n+1}^2 =\left[\left(v_{n_r}+v_{n_\theta}\cot\theta\right)^2+\frac{\ell^2}{r_n^2}\right]\tau_{n+1}^2+2r_n\left(v_{n_r}+v_{n_\theta}\cot\theta\right)\tau_{n+1}+r_n^2, \end{equation} and $v_{{n+1}_r}$ is given by~\eqref{eq:vel}. After evaluating $\sin\varphi_{n+1}$ and $\cos\varphi_{n+1}$ using~\eqref{eq:phi} and defining the convenient reduced variables $\rho\equiv r\sin\theta$, $u_r\equiv v_r\sin\theta,u_\theta\equiv v_\theta\cos\theta$, the mapping can be rewritten as (see Appendix) \begin{equation} \label{eq:reduced1} \rho_{n+1}^2 =\left[\left(u_{n_r}+u_{n_\theta}\right)^2+\frac{\ell^2}{\rho_n^2}\right]\tau_{n+1}^2+2\rho_n\left(u_{n_r}+u_{n_\theta}\right)\tau_{n+1}+\rho_n^2, \end{equation} \begin{multline} \label{eq:reduced2} u_{{n+1}_r} =\frac{\sin^2\theta}{\rho_{n+1}}\left\{\left[\left(u_{n_r}+u_{n_\theta}\right)^2+\frac{\ell^2}{\rho_n^2}\right]\tau_{n+1}+\rho_n\left(u_{n_r}+u_{n_\theta}\right)\right\}+u_{n_r}\cos^2\theta-u_{n_\theta}\sin^2\theta\\-\frac{1}{4}\tau_{n+1}\sin2\theta.\end{multline} Of course, our `reduced variable' $\rho$ is just the cylindrical polar coordinate; however, $u_r$ is not the corresponding velocity component $v_\rho$. In fact, $v_\rho=u_r+u_\theta$. Although this suggests that the conjugate variables $(\rho,v_\rho)$ should be used for the mapping, these two variables are \emph{insufficient} to determine the subsequent trajectory of the particle. This can be seen by writing energy conservation in terms of cylindrical polar coordinates: \begin{equation} 1=v_\rho^2+v_z^2+\frac{\ell^2}{\rho^2}+\rho\cot\theta. \end{equation} Thus if we know $\rho$ and $v_\rho$, the square of $v_z$ is given by \begin{equation} v_z^2=1-v_\rho^2-\frac{\ell^2}{\rho^2}-\rho\cot\theta.\end{equation}  However, without some additional piece of information regarding the sign of $v_z$, a priori we have no way of determining which square root to take. Thus $(r,v_r)$ or equivalently $(\rho,u_r)$ are the coordinates we use in this work. Although in principle we can eliminate $v_{\theta}$ or $u_\theta$ from the mapping via energy conservation, the equations become considerably more complicated and yield little additional insight. Moreover, we find that computing periodic orbits of the system is easier when using this `implicit' form of the mapping.\\\indent
The reason we choose $(r,v_r)$ as our primary mapping variables is simple: they are area-preserving i.e., the determinant of the Jacobian matrix ${\mathsf{J}}=\frac{\partial(r_{n+1},v_{{n+1}_r})}{\partial(r_n,v_{n_r})}$ is equal to unity. Another sensible choice would be $(\mathrm{sgn}(v_r) v_\parallel,v_\perp)$, where $v_\parallel$ and $v_\perp$ are the tangential and normal components of the velocity with respect to the cone boundary, respectively. Although these variables are sufficient to fully determine the dynamics, they are not area-preserving. In the next section, we shall utilize several times the area-preserving nature of $(r,v_r)$ in the computation of stability eigenvalues. In deriving specific fixed points and periodic orbits, the reduced mapping~\eqref{eq:reduced1}-\eqref{eq:reduced2} is more suitable. Since the variables are not area-preserving, however, they are not ideal for surface-of-section plots. 

\section{Analytical results\label{sec:3}}
In this section we examine some general properties of the mapping. In section~\ref{sec:3.1} we examine limiting cases of the system's parameter values, demonstrating that the conic billiard becomes integrable in several limits. In section~\ref{sec:3.2} we compute the system's fixed points and analyze their linear stability as a function of parameter values. 
\subsection{Simple properties of the mapping\label{sec:3.1}}
We note that if $\ell=0$ the conic system evidently reduces to the two-dimensional wedge billiard, as the Hamiltonian becomes \begin{equation} H_{\mathrm{wedge}}=\frac{1}{2}\left(p_r^2+\frac{p_\theta^2}{r^2}\right)+r\cos\theta. \end{equation} Hence, following \cite{miller86} we see that for $\ell=0$ the system becomes integrable for three cases (i) $\theta\rightarrow0^\circ$, (ii) $\theta=45^\circ$, and (iii) $\theta\rightarrow 90^\circ$. In the first case, the potential becomes purely radial in the limit i.e., \begin{equation}  H\rightarrow\frac{1}{2}\left(p_r^2+\frac{p_\theta^2}{r^2}\right)+r,\end{equation} implying the conservation of $p_\theta=r^2\dot{\theta}$, which is also preserved by the collision map.  For $\theta=45^\circ$, co-ordinates parallel and orthogonal to the wedge surface can be defined so that the motion becomes separable. Finally, for $\theta=90^\circ$ the motion becomes simply that of a projectile bounded from below by a horizontal floor. In this case, the motion is unbounded.\\\indent
For $\ell\neq0$, we lose integrability at $\theta=45^\circ$. However, we retain the integrable limit $\theta\rightarrow 90^\circ$, as the above argument is unaffected by the introduction of a nonzero $z$-component of angular momentum. Additionally, the limit $\theta\rightarrow0^\circ$ remains integrable as long as we simultaneously take $\ell\rightarrow\ell_{\mathrm{max}}$. Examining the Hamiltonian \begin{equation} H=\frac{1}{2}\left(p_r^2+\frac{p_\theta^2}{r^2}\right)+\frac{\ell^2}{r^2\sin^2\theta}+r\cos\theta, \end{equation} we see that if we naively take the limit $\theta\rightarrow0$ the term proportional to $\sin^{-2}\theta$ diverges. However, if we instead take $\theta\rightarrow0$ with $\ell\rightarrow\ell_{\mathrm{max}}\sim \tan\theta$, we see that \begin{equation} H\rightarrow\frac{1}{2}\left(p_r^2+\frac{p_\theta^2}{r^2}\right)+\frac{a}{r^2}+r,\end{equation} where $a=\frac{4}{27}$. Hence the potential again becomes indistinguishable from a central potential in this limit, implying that the quantity $p_\theta$ is conserved and the motion integrable.

\subsection{Fixed points\label{sec:3.2}}
\begin{figure}
\subfloat[]{\includegraphics[height=6.0cm, width=6.0cm]{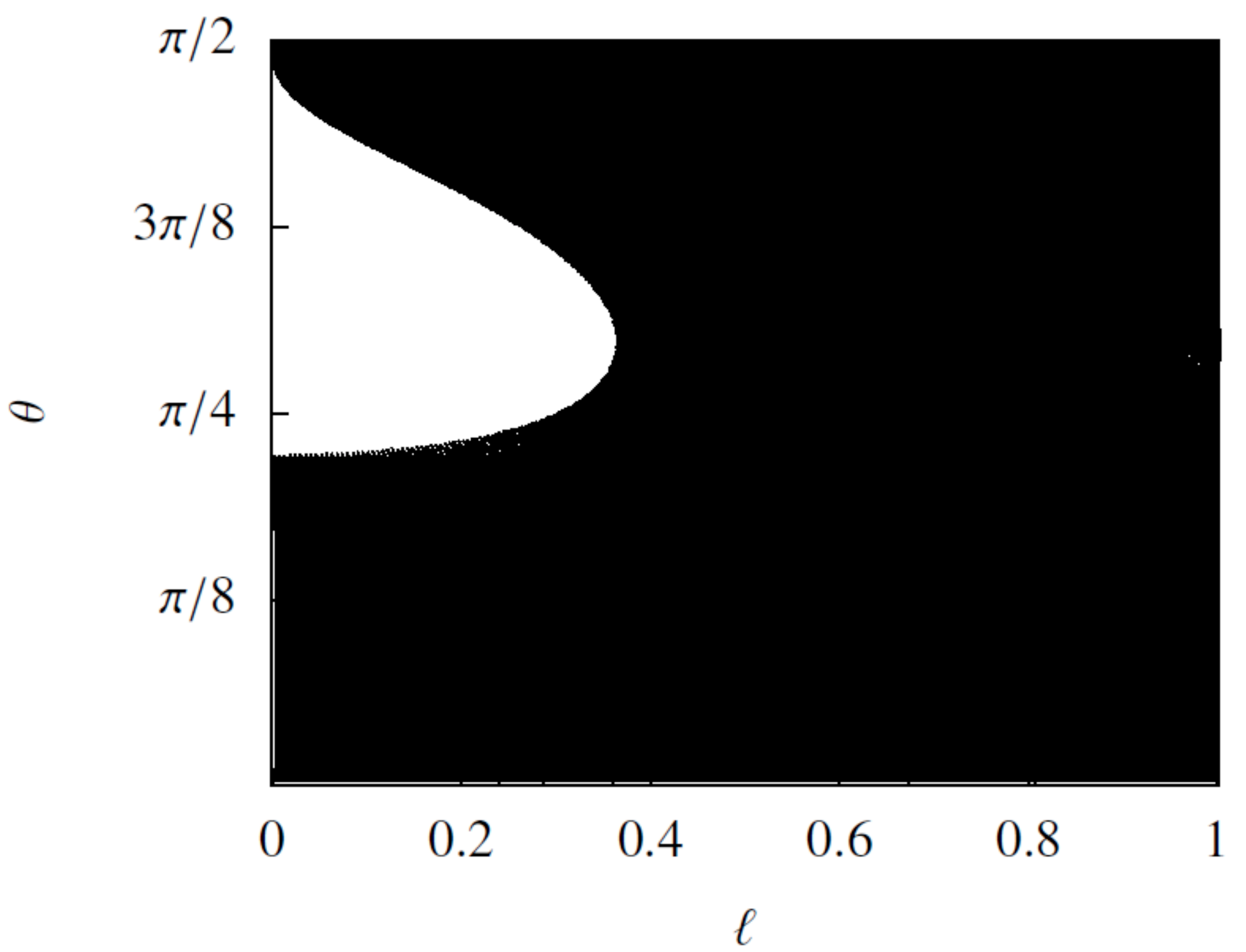}}
\label{fig:0a}
\qquad
\subfloat[]{\includegraphics[height=6.0cm, width=6.0cm]{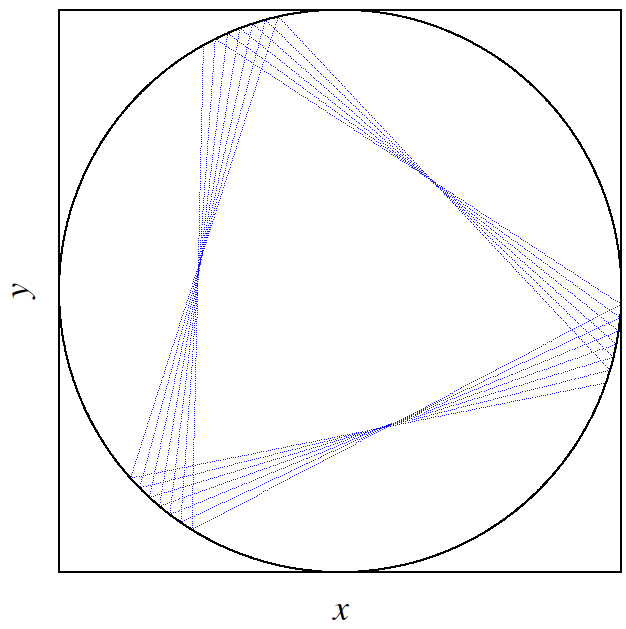}}
\label{fig:0c}
\caption{(a) Stability regions of the fixed point in the parameter $(\ell,\theta)$-space. Black regions correspond to an elliptic (stable) fixed point. Note that for $\ell=0$ the stability region is precisely $0\leqslant\theta\leqslant\frac{\pi}{4}$. (b) $xy$-projection of the real-time trajectory of the billiard for $\theta=\frac{\pi}{6}$ and $\ell=0.1$ for the first $20$ collisions of the fixed point.}\label{fig:0}
\end{figure}

Imposing the conditions $\rho_{n+1}=\rho_n\equiv\rho$, $u_{{n+1}_r}=u_{n_r}\equiv u_r$ on the mapping, we arrive at (here $\tau_{n+1}=\tau=\mathrm{const.}$) \begin{equation} 0=\left[(u_r+u_\theta)^2+\frac{\ell^2}{\rho^2}\right]\tau+2\rho(u_r+u_\theta),\end{equation} \begin{equation} u_r=\frac{\sin^2\theta}{\rho}\left\{\left[(u_r+u_\theta)^2+\frac{\ell^2}{\rho^2}\right]\tau+\rho(u_r+u_\theta)\right\}+u_r\cos^2\theta-u_\theta\sin^2\theta-\frac{\tau}{4}\sin2\theta.\end{equation} The first equation is equivalent to \begin{equation} \tau=\frac{-2\rho(u_r+u_\theta)}{(u_r+u_\theta)^2+\ell^2/\rho^2} .\end{equation} Evidently this holds for all fixed points of the map. Using this in the mapping equations for $u_{{n+1}_r}$ and $u_{{n+1}_\theta}$ gives the simultaneous equations \begin{equation} \tau=-4(u_r+u_\theta)\tan\theta\qquad\qquad\mathrm{and}\qquad\qquad \tau=-4u_r\cot\theta-4u_\theta\tan\theta. \end{equation} Thus either $u_r=0$ or $\theta=\frac{\pi}{4}$. In the case $u_r=0$, setting the two expressions for $\tau$ equal gives \begin{equation} \label{eq:fixed1}\rho=2\tan\theta\left(u_\theta^2+\frac{\ell^2}{\rho^2}\right). \end{equation} If $\ell=0$, then it is easily verified that \begin{equation}\label{eq:ellzero}  \rho=\frac{\sin2\theta}{2+\cos2\theta},\qquad u_r=0\end{equation} is a fixed point; this agrees with the fixed point of the wedge billiard~\cite{miller86}. In the general case that $\ell\neq0$ we can apply energy conservation to arrive at \begin{equation}\label{eq:fixed2} u_\theta^2=\frac{(\rho^2-3\ell^2)\cos^2\theta}{\rho^2(2+\cos2\theta)},\end{equation}  which must hold simultaneously with~\eqref{eq:fixed1}. Eliminating $u_\theta$ in~\eqref{eq:fixed1} and~\eqref{eq:fixed2} yields the cubic equation 
\begin{equation} \left[(2+\cos2\theta)\cot\theta\right]\rho^3-\left(2\cos^2\theta\right)\rho^2-2\ell^2\sin^2\theta=0.  \label{eq:fixedpointsoln}\end{equation} Clearly, if we set $\ell=0$ in the above equations we recover the limiting case found above. Although the equation determining $\rho$ is cubic, it turns out that for all allowable parameter values only one of the three roots is real. Hence there is a single fixed point \begin{equation} \rho=\rho_*,\qquad u_r=0,\qquad \mathrm{where}\,\,\,\rho_*\,\,\,\mathrm{solves}\,\,\,\eqref{eq:fixedpointsoln}. \end{equation} The stability of this orbit can be analyzed in terms of the eigenvalues of the Jacobian matrix. As is shown in e.g., \cite{berry1981}, the stability condition that the eigenvalues of the Jacobian are less than one in modulus is equivalent to, for an area-preserving map, the condition \begin{equation} \vert\mathrm{Tr}\,{\mathsf{J}}\vert<2. \end{equation} This can be alternatively stated in terms of \emph{Green's residue} \cite{green1981}, defined as \begin{equation} R=\frac{2-\mathrm{Tr}\,{\mathsf{J}}}{4}.\end{equation} In terms of $R$, a fixed point is stable for $0<R<1$, and unstable for $R<0$ and $R>1$. Stable fixed points are also called elliptic, and unstable fixed points are called hyperbolic. Of course, since the $(\rho,u_r)$ map is not area-preserving, we must use the Jacobian of the $(r,v_r)$ map in stability calculations. In terms of these variables the fixed point is \begin{equation} r=r_*,\qquad\qquad v_r=0,\end{equation} where $r_*$ is given by \begin{equation} \left[(2+\cos2\theta)\cos\theta\right] r_*^3-\left(2\cos^2\theta\right) r_*^2-2\ell^2=0.\end{equation} The real root of this cubic can be written \begin{equation} r_*=k(\ell,\theta)+\frac{4\cos^2\theta}{9(2+\cos2\theta)^2 k(\ell,\theta)}+\frac{2\cos\theta}{3(2+\cos2\theta)}, \end{equation} where $k(\ell,\theta)$ is \begin{equation} k(\ell,\theta)\equiv\left[\frac{8\cos^3\theta}{27(2+\cos2\theta)^3}+\frac{\ell^2\sec\theta}{2+\cos2\theta}+\frac{\ell\sec\theta}{3\sqrt{3}(2+\cos2\theta)^2}\sqrt{16\cos^4\theta+27\ell^2(2+\cos2\theta)^2}\right]^{1/3}.\end{equation} It is easy to see that this agrees with the case $\ell=0$ in equation~\eqref{eq:ellzero}. Since the mapping depends \emph{explicitly} on the time between collisions $\tau_{n+1}$, in computing the necessary derivatives we must take care to use the chain rule i.e., \begin{equation} \frac{\partial r_{n+1}}{\partial r_n}=\frac{\partial r_{n+1}}{\partial r_n}\Big\vert_{\tau_{n+1}=\mathrm{const.}}+\frac{\partial r_{n+1}}{\partial \tau_{n+1}}\frac{\partial \tau_{n+1}}{\partial r_n}, \qquad  \frac{\partial v_{{n+1}_r}}{\partial v_{n_r}}=\frac{\partial v_{{n+1}_r}}{\partial v_{n_r}}\Big\vert_{\tau_{n+1}=\mathrm{const.}}+\frac{\partial v_{{n+1}_r}}{\partial \tau_{n+1}}\frac{\partial \tau_{n+1}}{\partial v_{n_r}}, \end{equation} where the derivatives of $\tau_{n+1}$ are computed using implicit differentiation on the cubic equation for the time interval i.e., \begin{equation}\frac{\partial \tau_{n+1}}{\partial r_n}=-\frac{\partial f/\partial r_n}{\partial f/\partial \tau_{n+1}}, \end{equation} where $f(\tau_{n+1},r_n,v_{n_r})=0$ is defined by the left hand side of~\eqref{eq:time}. The computations are straightforward but tedious due to the fact that $v_{n_\theta}$ depends on $r_n$ and $v_{n_r}$ in the energy expression~\eqref{eq:energy}. The explicit expression for $\mathrm{Tr}\,{\mathsf{J}}$, or equivalently Green's residue $R$ is long and complicated, and will not be included here. Instead, in Fig.~\ref{fig:0}(a) we display the stability regions in $(\ell,\theta)$-space.

We note that although $(r_*,0)$ is a fixed point of the \emph{mapping}, the physical trajectory of the billiard does not in general form a closed orbit. Instead, fixing $r=r_*$ at each collision (which is equivalent to fixing the height of the particle at each collision) constrains the collision points to a circle on the conic boundary surface. It is clear that the trajectory of the billiard will form a closed path if and only if (note that $\varphi_{n+1}=\varphi=\mathrm{const.}$ for a fixed point)\begin{equation} \varphi=\frac{2\pi p}{q},\quad \mathrm{where}\quad p,q\in\mathbb{Z}^+. \end{equation} Physically, this orbit corresponds to a trajectory which repeats after $q$ collisions, during which the billiard's $\phi$-coordinate cycles through $2\pi p$ radians. When $\varphi$ is not a rational multiple of $2\pi$, the physical trajectory is quasiperiodic. Since the fixed point orbit is confined to constant $z$ and $v_z$, the physical trajectory corresponding to this orbit is equivalent (when viewed from the positive $z$-axis) to that of the \emph{circular billiard} i.e., a billiard moving uniformly within a circle. In Fig.~\ref{fig:0}(b) we plot the projection of the physical trajectory in the $xy$-plane for $\theta=30^\circ$ or $\frac{\pi}{6}$ radians and $\ell=0.1$.

For more general periodic orbits, we note that for $\ell=0$ there are, for example, well-known \cite{miller86} period-$m$ orbits of the form \begin{equation} \rho=\tan\theta\left[1-\frac{m^2\cos^2\theta+\sin^2\theta}{1-(m+1)^2\cos^2\theta\cos2\theta}\right],\qquad u_r=\frac{m\sin\theta\cos\theta}{\sqrt{1-(m+1)^2\cos^2\theta\cos2\theta}}.\end{equation} Such solutions clearly hold for the conic billiard when $\ell=0$; however, generalizing these solutions to $\ell\neq0$ is difficult to do analytically due to the implicit nature of the mapping. However, the existence of higher order periodic orbits can be confirmed numerically. 

\section{Numerical results\label{sec:4}}
In this section, we iterate the Poincar\'{e} map numerically to examine the qualitative changes in the phase space as the parameters of the system are varied for two values of $\ell$ which are representative of the dynamical behavior. In section~\ref{sec:4.1} we consider a relatively small value of $\ell=0.1$, which corresponds to small values of the particle's $z$-component of angular momentum. Intuitively, we expect the behavior of the system in this case to be qualitatively similar to that of the wedge billiard \cite{miller86,richter1990}. In section~\ref{sec:4.2}, we increase the particle's angular momentum to $\ell=0.5$ and examine how the wedge-like behavior is destroyed and, in the limit $\ell\rightarrow1$, approaches integrability.

\subsection{Wedge-like behavior: small $\ell$\label{sec:4.1}}
For $\ell=0$ it is clear that the conic billiard reduces to the wedge billiard. For relatively small values of $0<\ell\lesssim0.2$ we find that some structures found in the wedge system survive; however, the correspondence is not exact. Specifically, the conic billiard does not in general become ergodic for angles above an exact value ($45^\circ$, for the wedge). Instead, for small $\ell$ values the conic system becomes ergodic in only a certain range of angles, with stable periodic orbits reappearing for large angles.

\begin{figure}
\subfloat[]{\includegraphics[height=4.5cm, width=4.4cm]{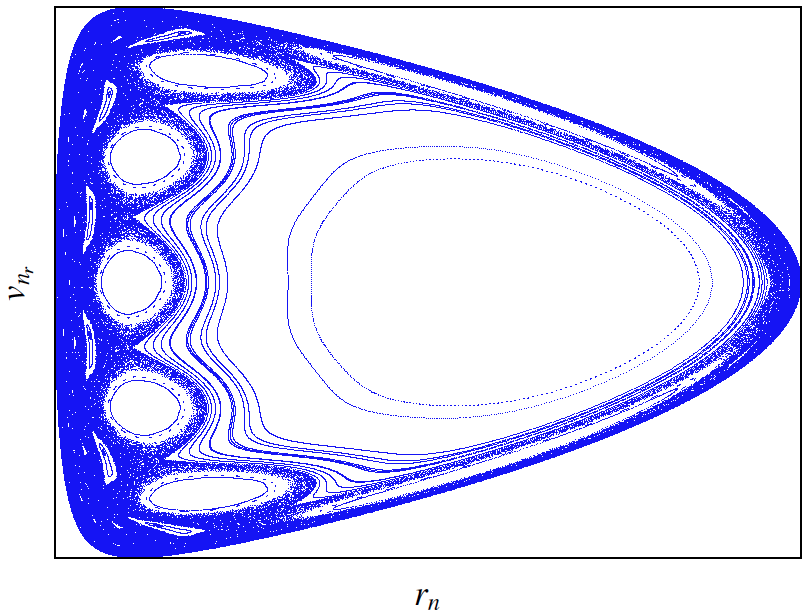}}
\label{fig:1a}
\qquad
\subfloat[]{\includegraphics[height=4.5cm, width=4.4cm]{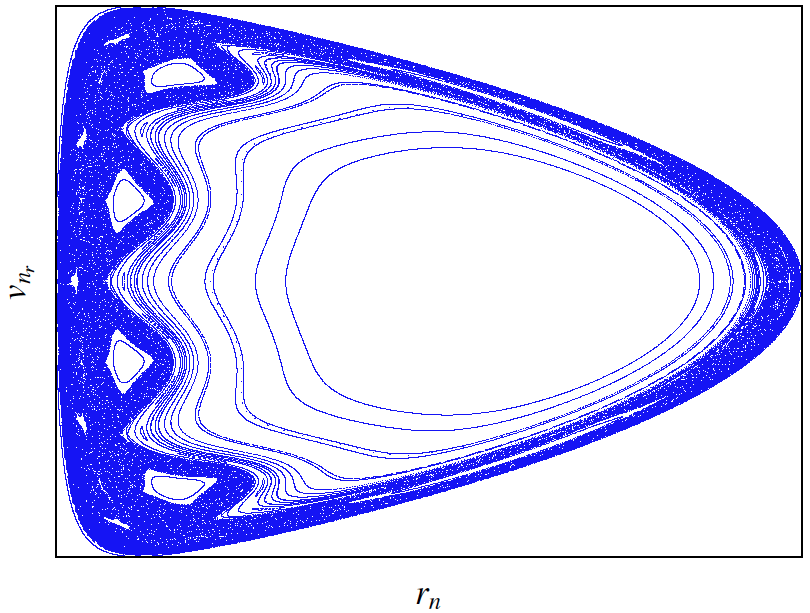}}
\label{fig:1b}
\qquad
\subfloat[]{\includegraphics[height=4.5cm, width=4.4cm]{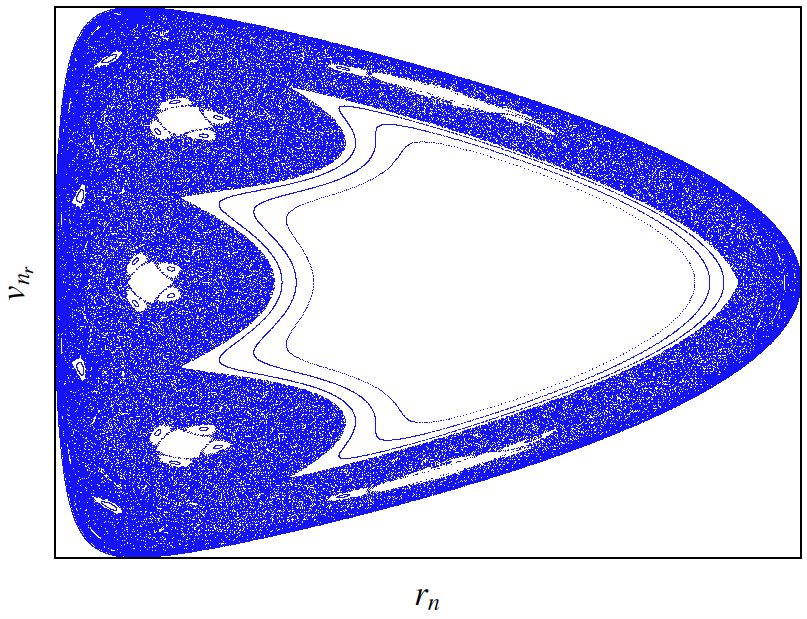}}\label{fig:1c}
\qquad 
\subfloat[]{\includegraphics[height=4.5cm, width=4.4cm]{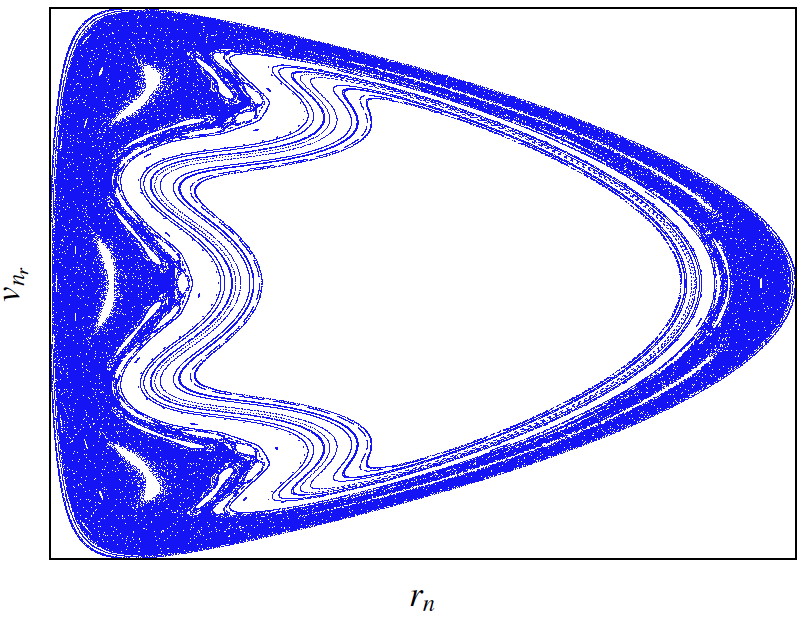}}\label{fig:1d}
\qquad
\subfloat[]{\includegraphics[height=4.5cm, width=4.4cm]{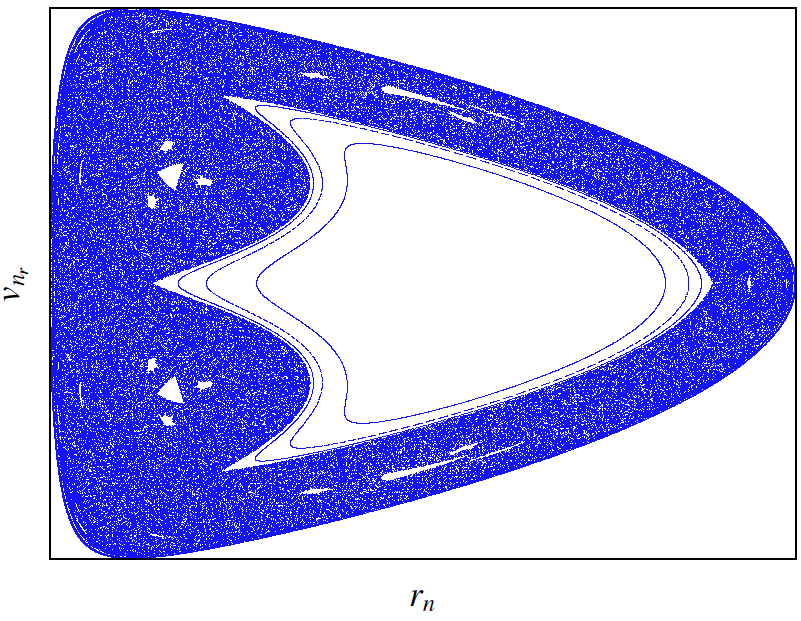}}\label{fig:1e}
\qquad
\subfloat[]{\includegraphics[height=4.5cm, width=4.4cm]{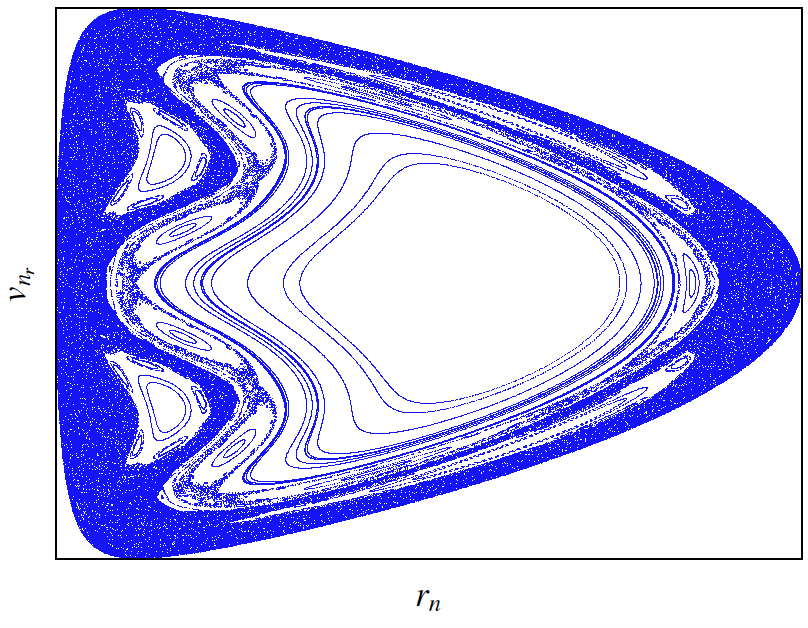}}
\label{fig:1f}
\caption{SOS at $\ell=0.1$ for (a) $\theta=15^\circ$; (b) $\theta=18.5^\circ$; (c) $\theta=21^\circ$; (d) $\theta=24.5^\circ$; (e) $\theta=27^\circ$; (f) $\theta=30.5^\circ$.}
\label{fig:small_ell1}
\end{figure}

\begin{figure}
\subfloat[]{\includegraphics[height=4.5cm, width=4.4cm]{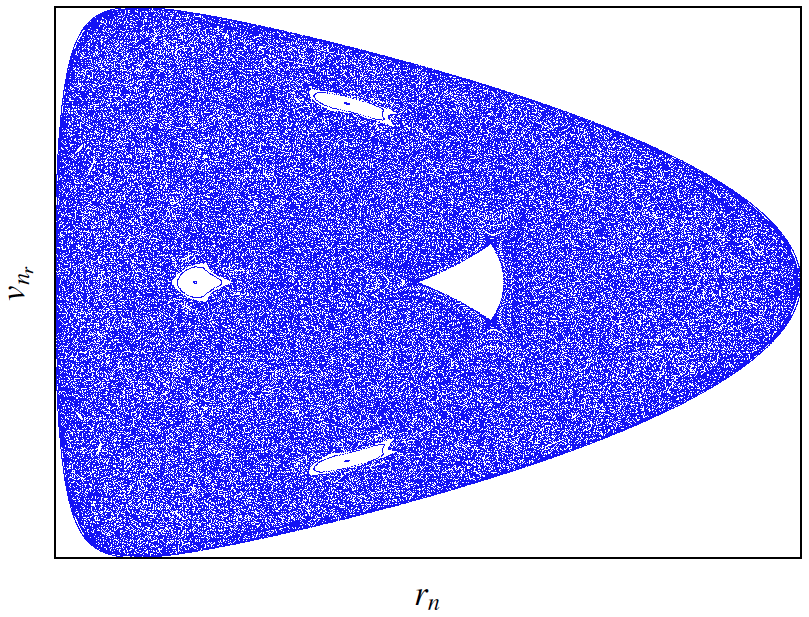}}\label{fig:2a}
\qquad
\subfloat[]{\includegraphics[height=4.5cm, width=4.4cm]{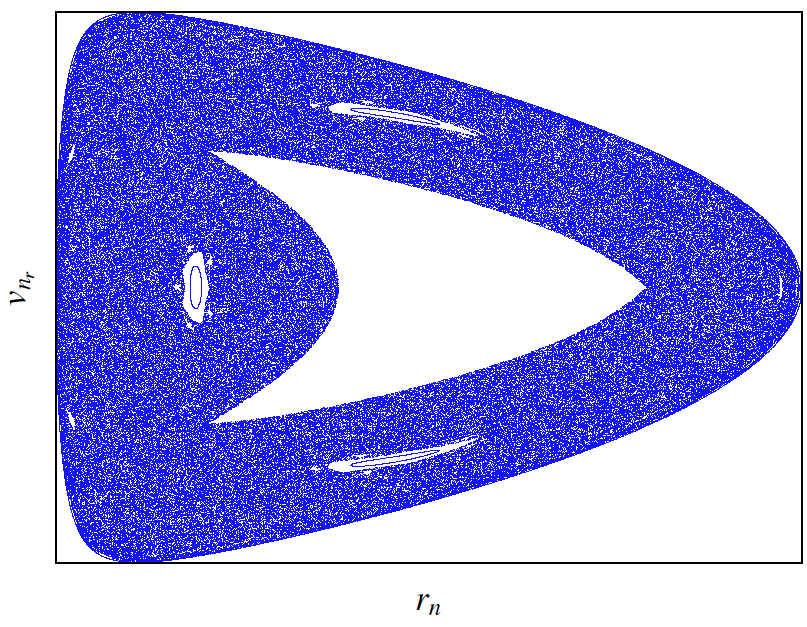}}\label{fig:2b}
\qquad
\subfloat[]{\includegraphics[height=4.5cm, width=4.4cm]{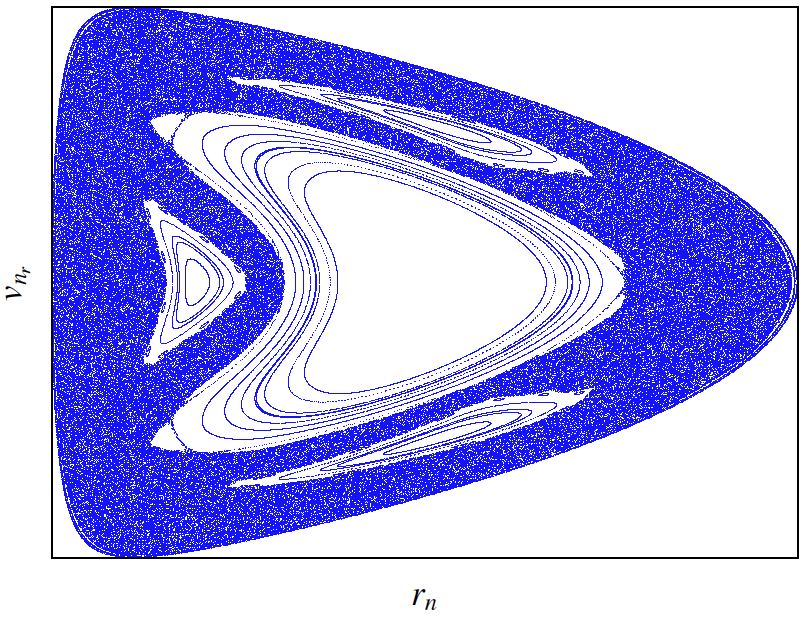}}\label{fig:2c}
\qquad
\subfloat[]{\includegraphics[height=4.5cm, width=4.4cm]{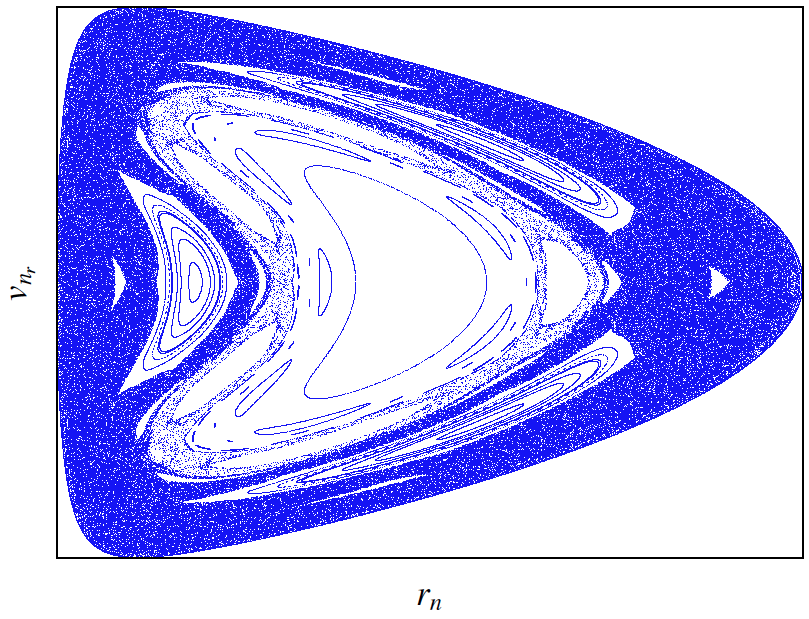}}\label{fig:2d}
\qquad
\subfloat[]{\includegraphics[height=4.5cm, width=4.4cm]{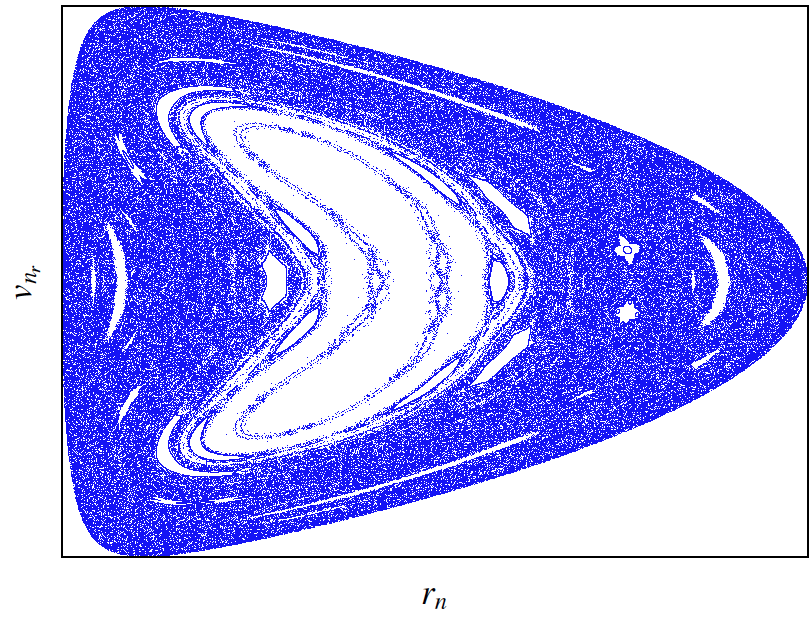}}\label{fig:2e}
\qquad 
\subfloat[]{\includegraphics[height=4.5cm, width=4.4cm]{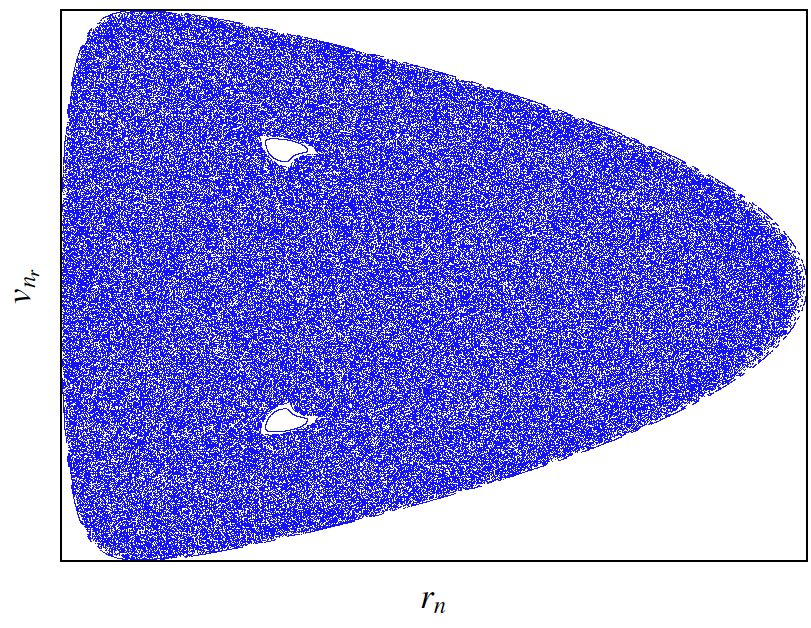}}\label{fig:2f}
\caption{SOS at $\ell=0.1$ for (a) $\theta=34^\circ$; (b) $\theta=37.5^\circ$; (c) $\theta=41^\circ$; (d) $\theta=44.5^\circ$; (e) $\theta=47^\circ$; (f) $\theta=50.5^\circ$.}
\label{fig:small_ell2}
\end{figure}

\begin{figure}
\subfloat[]{\includegraphics[height=4.5cm, width=4.4cm]{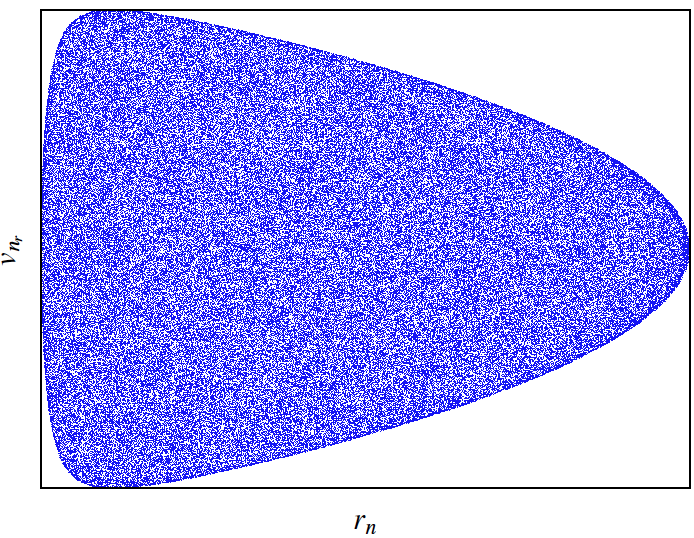}}\label{fig:3a}
\qquad
\subfloat[]{\includegraphics[height=4.5cm, width=4.4cm]{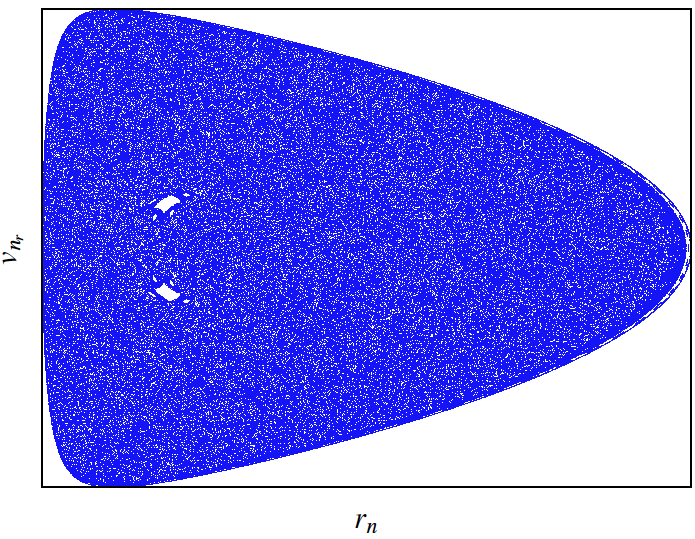}}\label{fig:3b}
\qquad
\subfloat[]{\includegraphics[height=4.5cm, width=4.4cm]{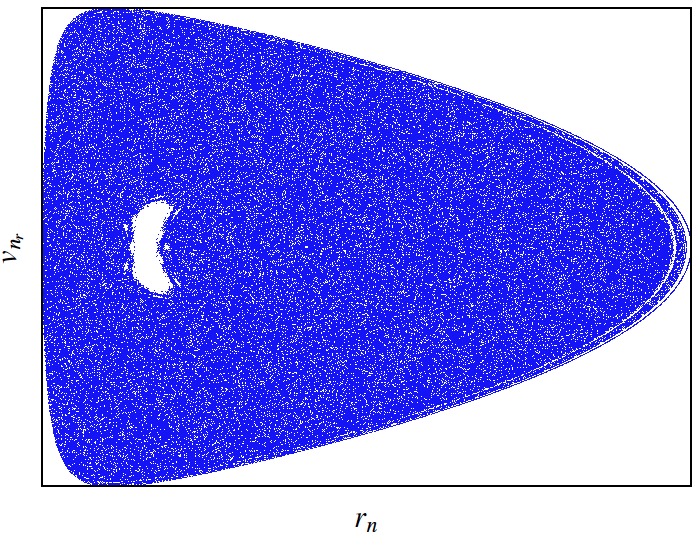}}\label{fig:3c}
\qquad
\subfloat[]{\includegraphics[height=4.5cm, width=4.4cm]{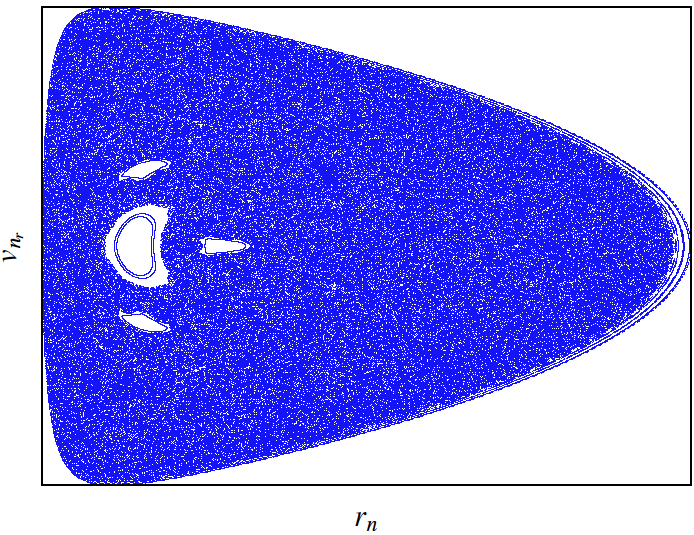}}\label{fig:3d}
\qquad
\subfloat[]{\includegraphics[height=4.5cm, width=4.4cm]{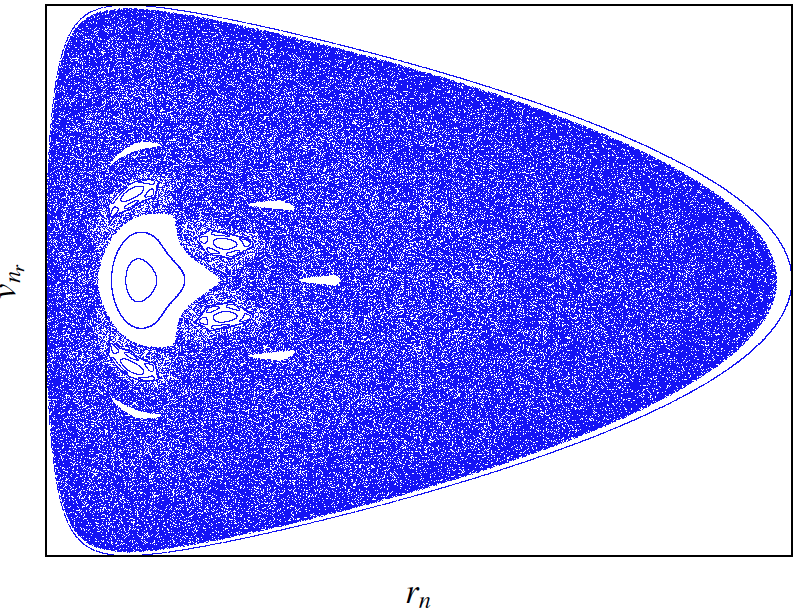}}\label{fig:3e}
\qquad 
\subfloat[]{\includegraphics[height=4.5cm, width=4.4cm]{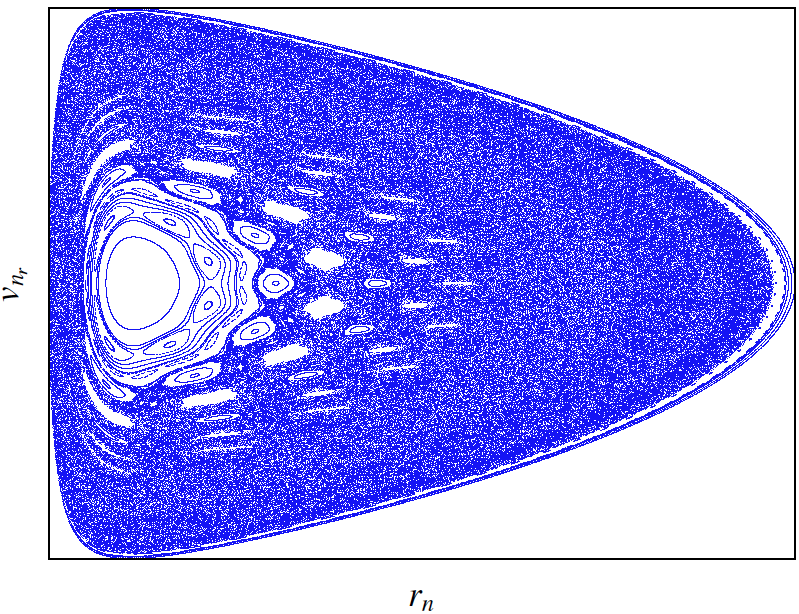}}\label{fig:3f}
\qquad
\subfloat[]{\includegraphics[height=4.5cm, width=4.4cm]{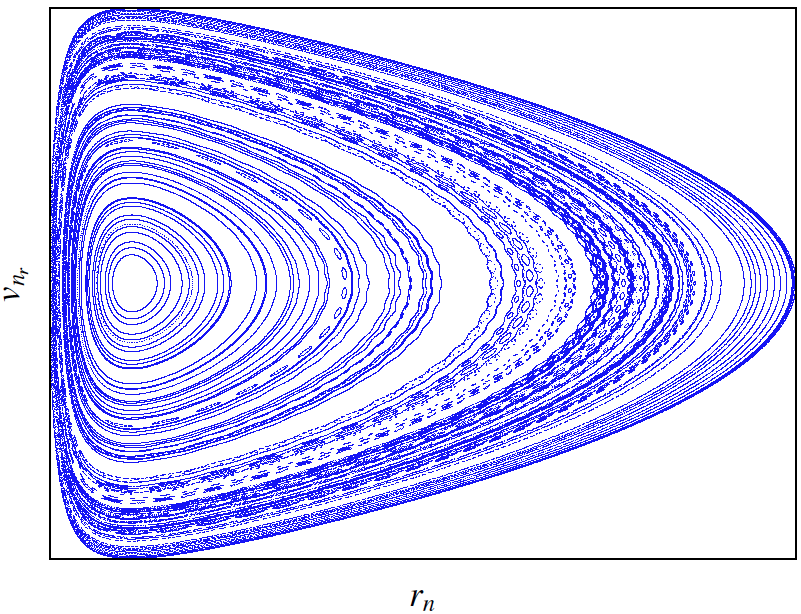}}\label{fig:3g}
\qquad
\subfloat[]{\includegraphics[height=4.5cm, width=4.4cm]{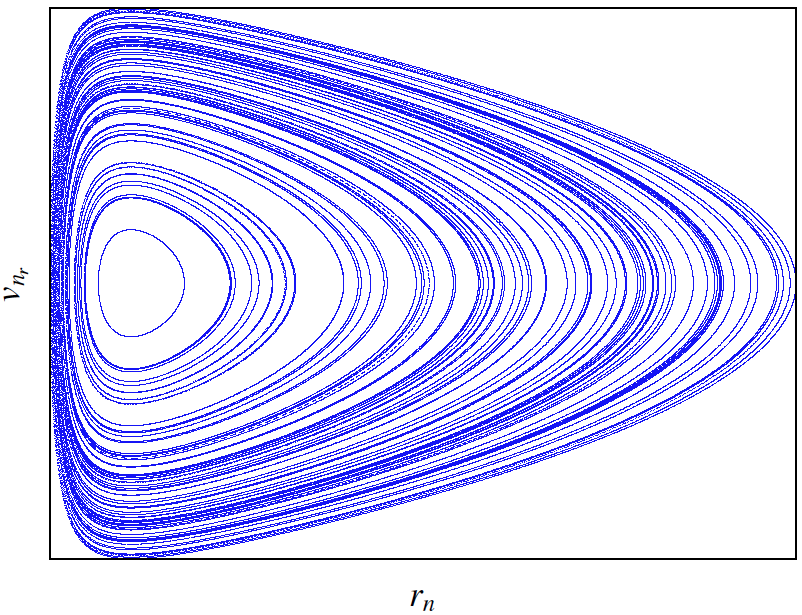}}\label{fig:3h}
\caption{SOS at $\ell=0.1$ for (a) $\theta=54^\circ$; (b) $\theta=73^\circ$; (c) $\theta=74.5^\circ$; (d) $\theta=77^\circ$; (e) $\theta=80.5^\circ$; (f) $\theta=84^\circ$; (g) $\theta=87.5^\circ$; (h) $\theta=89.5^\circ$.}
\label{fig:small_ell3}
\end{figure}

In Figs.~\ref{fig:small_ell1}-\ref{fig:small_ell3} we display the Poincar\'{e} surface-of-section $(r_n,v_{n_r})$ for a variety of cone angles at $\ell=0.1$. We note that the general structure of the phase space is quite typical of two-dimensional area-preserving maps. As $\theta$ is varied, we see a number of interesting bifurcation processes as periodic orbits lose and gain stability. The fixed point found in section~\ref{sec:3.2}, together with its surrounding KAM islands, is seen in Fig.~\ref{fig:small_ell1}(a)-(f), which corresponds to $15^\circ\lesssim\theta\lesssim30^\circ$ to constitute the majority of the phase space along with chaotic regions. Additionally, for this range of angles there are a variety of period-$2$,$3$ and higher orbits with their own surrounding island structures. Note that in these figures the elliptic nature of the fixed point is clear, which agrees with the stability calculations of section~\ref{sec:3.2}. In Fig.~\ref{fig:small_ell2}(a) at $\theta=34^\circ$ the fixed point becomes hyperbolic, and aside from a single period-$3$ orbit the majority of the phase space becomes chaotic. As $\theta$ is increased further to $\theta\simeq45^\circ$ the stability regions grow, leading to bifurcations and higher periodic orbits, as seen in Fig.~\ref{fig:small_ell2}(d)-(e). Finally, in Fig.~\ref{fig:small_ell2}(f) all periodic orbits disappear except for a single period-$2$ orbit, which remains until $\theta\simeq54^\circ$, where the system appears to be ergodic. This behavior is analogous to that of the wedge.

The apparent ergodicity remains for $54^\circ\lesssim\theta\lesssim73^\circ$, where a stable period-$2$ orbit re-appears, as shown in Fig.~\ref{fig:small_ell3}(b). As $\theta$ is increased to $\theta\simeq74.5^\circ$, the stability islands collapse to the single fixed point of Fig.~\ref{fig:small_ell3}(c). As seen in Fig.~\ref{fig:small_ell3}(d)-(f), higher periodic orbits gradually appear and the stability region of the fixed point grows as the angle of the cone is increased further. Finally, for large cone angles the system approaches integrability, with the chaotic part of the phase space vanishing in the limit $\theta\rightarrow90^\circ$. 

\subsection{Large $\ell$ values\label{sec:4.2}}
\begin{figure}
\subfloat[]{\includegraphics[height=4.5cm, width=4.4cm]{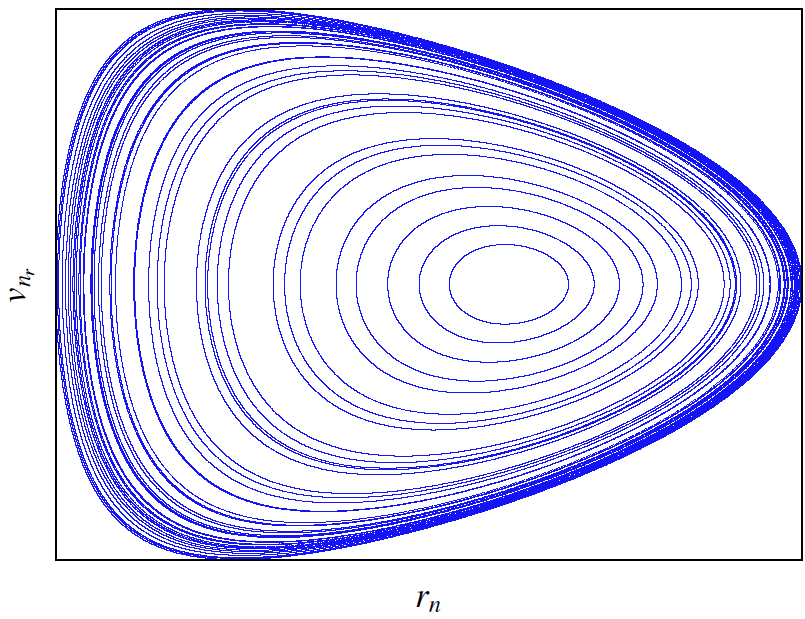}}\label{fig:4a}
\qquad
\subfloat[]{\includegraphics[height=4.5cm, width=4.4cm]{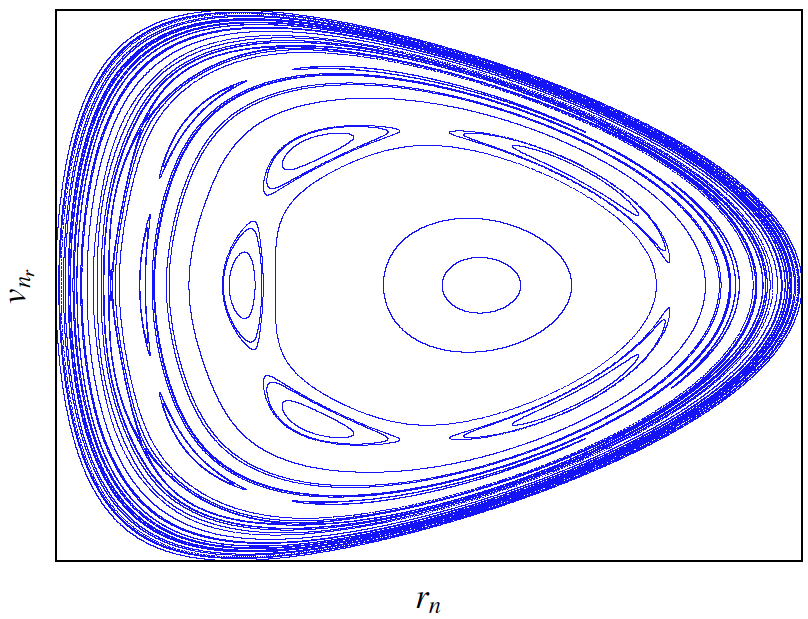}}\label{fig:4b}
\qquad
\subfloat[]{\includegraphics[height=4.5cm, width=4.4cm]{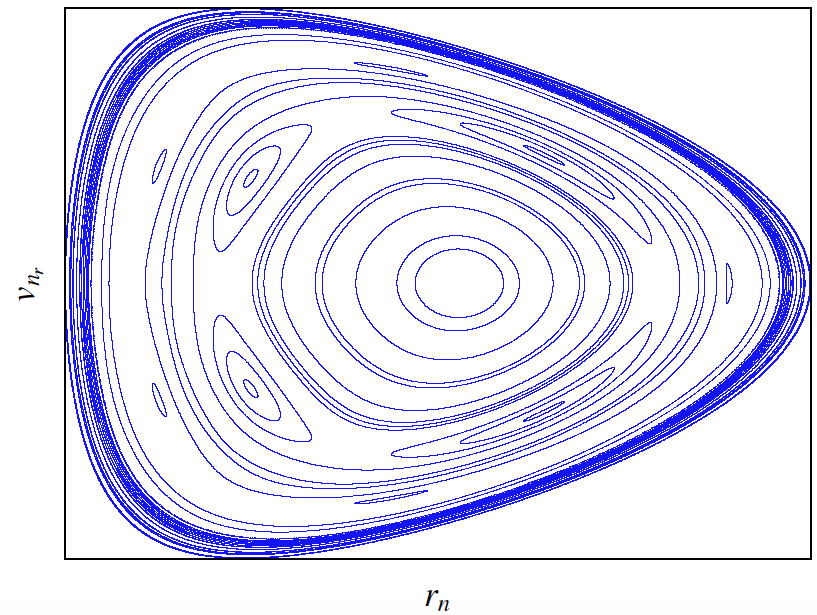}}\label{fig:4c}
\qquad 
\subfloat[]{\includegraphics[height=4.5cm, width=4.4cm]{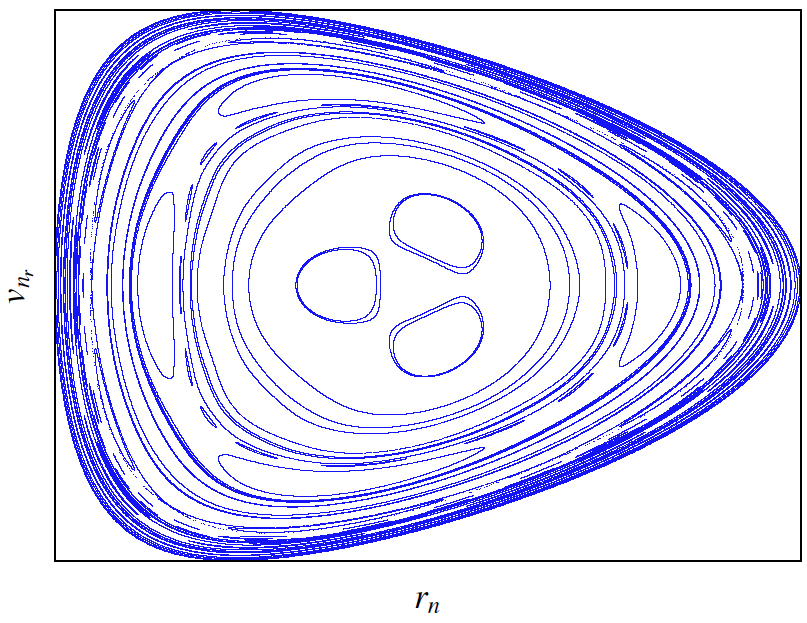}}\label{fig:4d}
\qquad
\subfloat[]{\includegraphics[height=4.5cm, width=4.4cm]{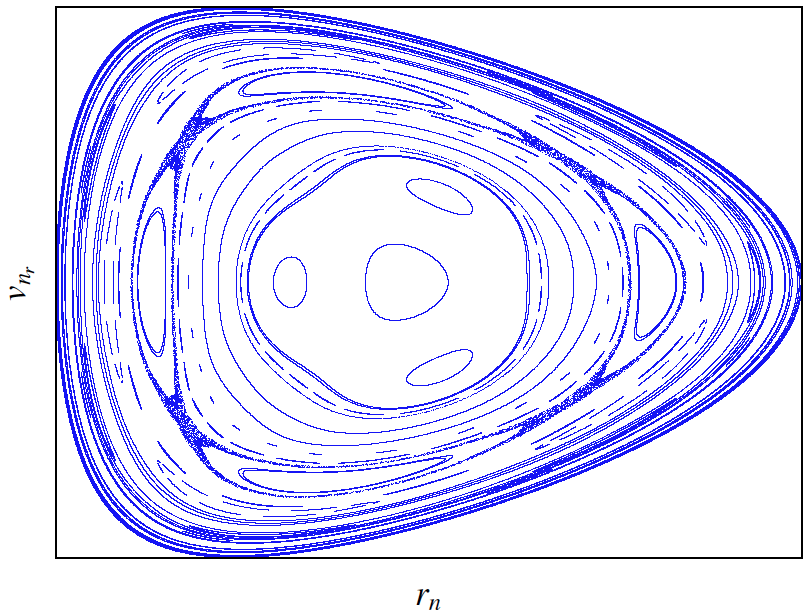}}\label{fig:4e}
\qquad
\subfloat[]{\includegraphics[height=4.5cm, width=4.4cm]{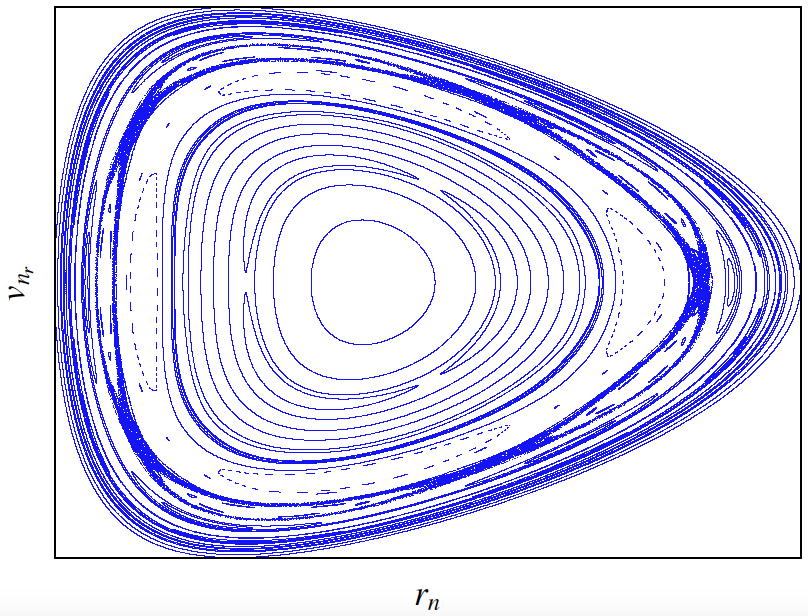}}\label{fig:4f}
\caption{SOS at $\ell=0.5$ for (a) $\theta=10^\circ$; (b) $\theta=25.5^\circ$; (c) $\theta=34^\circ$; (d) $\theta=42^\circ$; (e) $\theta=44^\circ$; (f) $\theta=50^\circ$.}
\label{fig:large_ell1}
\end{figure}

\begin{figure}
\subfloat[]{\includegraphics[height=4.5cm, width=4.4cm]{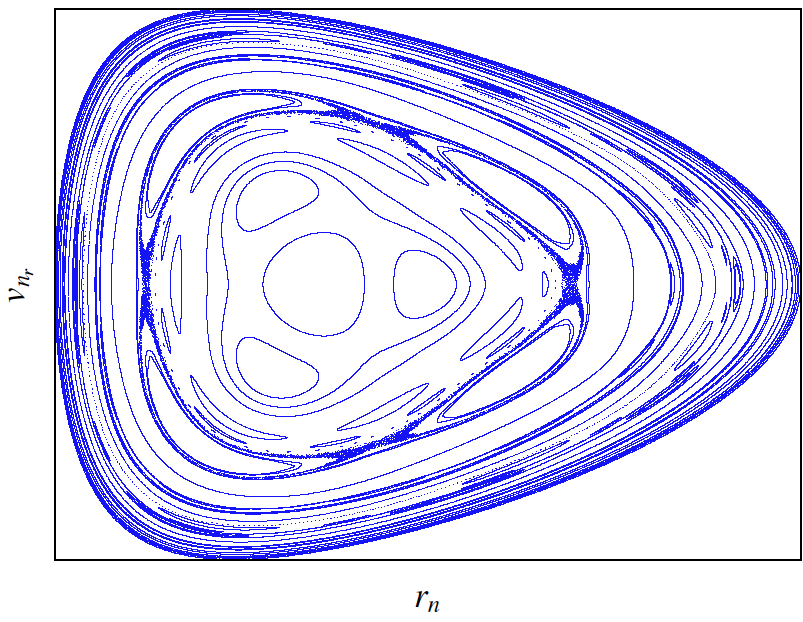}}\label{fig:5a}
\qquad
\subfloat[]{\includegraphics[height=4.5cm, width=4.4cm]{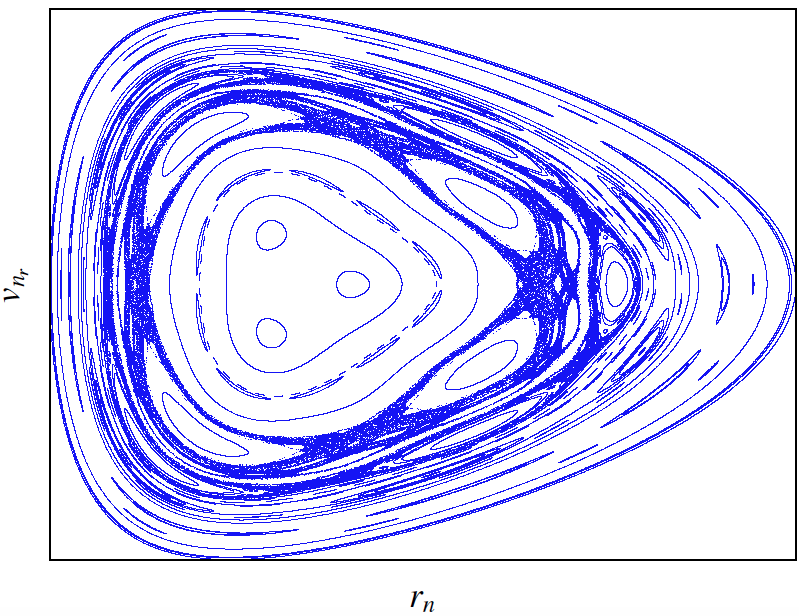}}\label{fig:5b}
\qquad
\subfloat[]{\includegraphics[height=4.5cm, width=4.4cm]{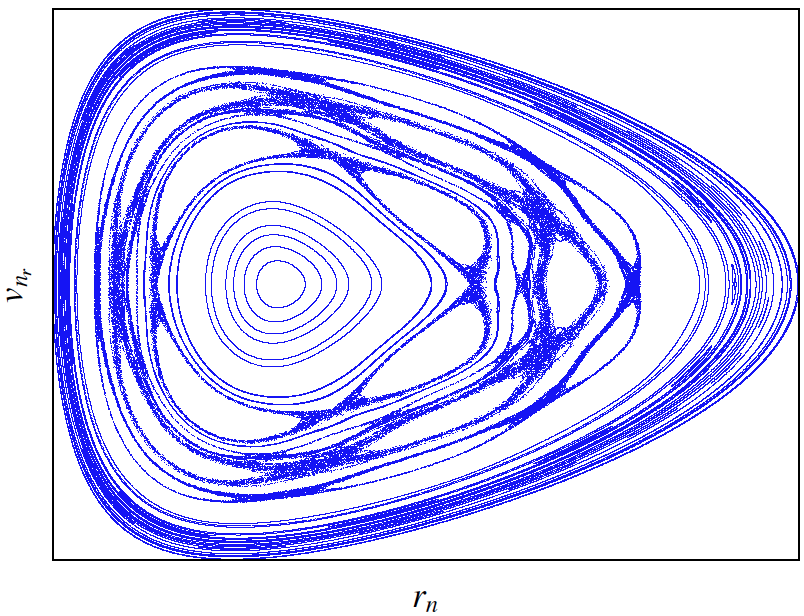}}\label{fig:5c}
\qquad
\subfloat[]{\includegraphics[height=4.5cm, width=4.4cm]{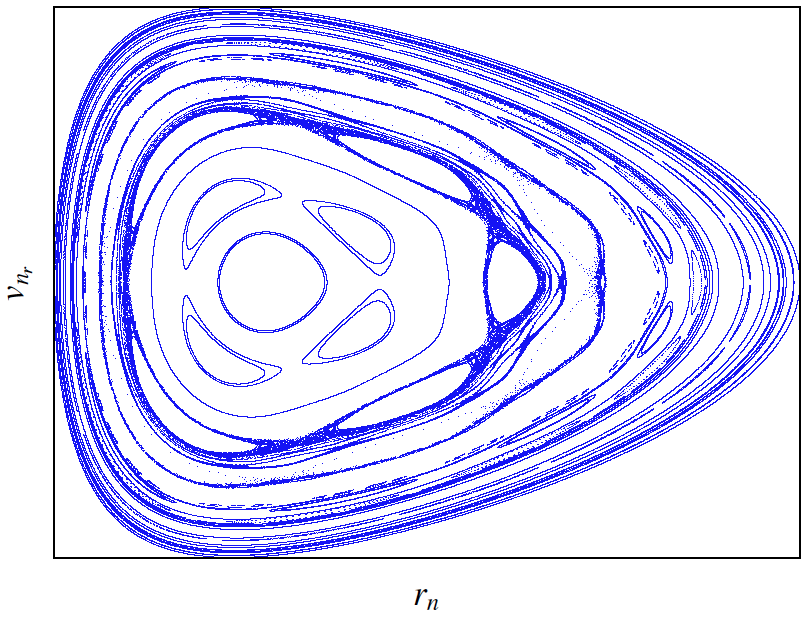}}\label{fig:5d}
\qquad
\subfloat[]{\includegraphics[height=4.5cm, width=4.4cm]{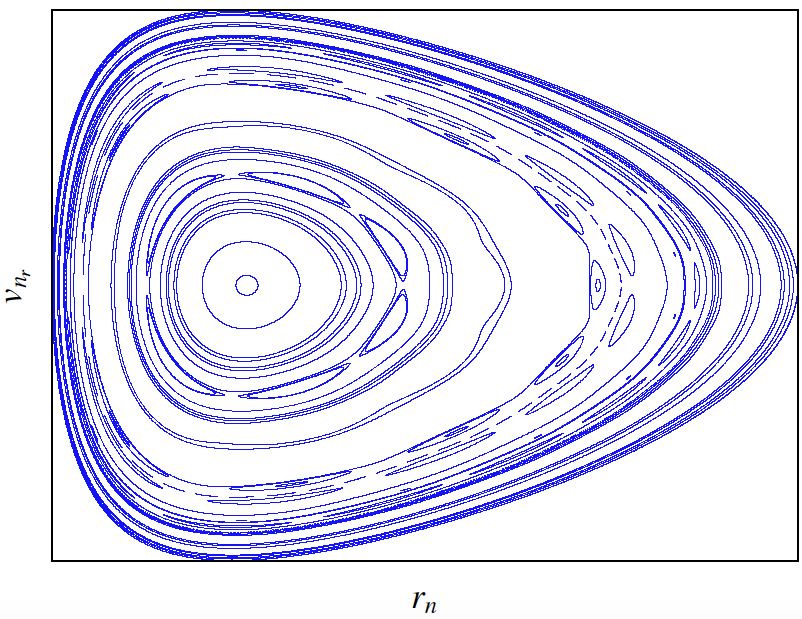}}\label{fig:5e}
\qquad 
\subfloat[]{\includegraphics[height=4.5cm, width=4.4cm]{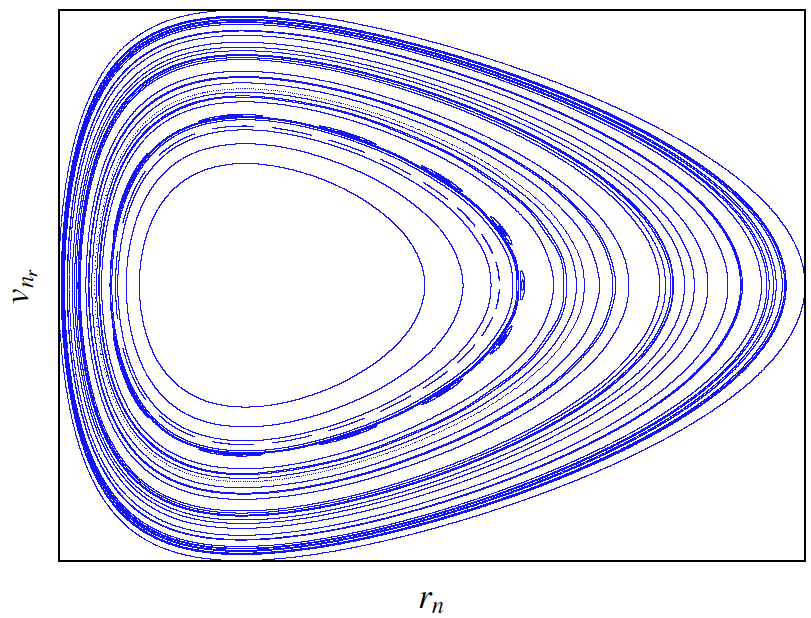}}\label{fig:5f}
\caption{SOS at $\ell=0.5$ for (a) $\theta=60^\circ$; (b) $\theta=63.5^\circ$; (c)  $\theta=67^\circ$; (d) $\theta=70.5^\circ$; (e) $\theta=77^\circ$; (f) $\theta=81.5^\circ$.}
\label{fig:large_ell2}
\end{figure}

As $\ell$ is increased, the relative amount of chaos in the phase space becomes smaller. As we increase the allowable $v_\phi$, we lose the ``wedge-like'' behavior seen in Figs.~\ref{fig:small_ell1}-\ref{fig:small_ell3} and the dynamics become more regular. In Fig.~\ref{fig:large_ell1} and Fig.~\ref{fig:large_ell2} we plot the Poincar\'{e} surface-of-section $(r_n,v_{n_r})$ for $\ell=0.5$ and a variety of $\theta$ values. The value $\ell=0.5$ is representative of the behavior of the system in the range $0.1<\ell\lesssim0.7$. As $\ell$ is increased beyond this range, the behavior approaches the integrable limit $\ell\rightarrow1$, with the amount of chaos becoming negligible (and disappearing completely in the limit). Additionally, we find no instances of a simply connected region of chaos.

For $0\leqslant\theta\lesssim10^\circ$ the system's behavior is nearly integrable, as may be seen in Fig.~\ref{fig:large_ell1}(a). The fixed point found in section~\ref{sec:3.2} is once again seen to play a significant role in the phase space, with the chaotic region remaining quite small until $\theta\simeq42^\circ$. As $\theta$ is increased further, a period-$4$ orbit shown in Fig.~\ref{fig:large_ell1}(d) collides with a stable island surrounding the fixed point, leading to the chaotic band seen in Fig.~\ref{fig:large_ell1}(e) at $\theta=44^\circ$. This chaotic band remains and reaches its maximum size for $\theta\simeq63.5^\circ$, as seen in Fig.~\ref{fig:large_ell2}(b). For larger cone angles the chaotic band gradually shrinks, vanishing in the integrable limit $\theta\rightarrow90^\circ$. 

\begin{figure}
\subfloat[]{\includegraphics[height=4.5cm, width=4.4cm]{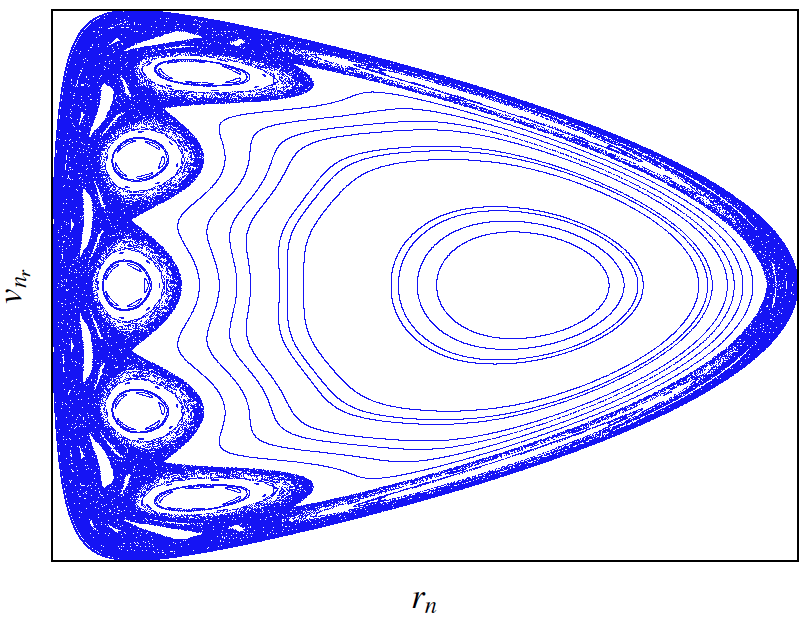}}\label{fig:6a}
\qquad
\subfloat[]{\includegraphics[height=4.5cm, width=4.4cm]{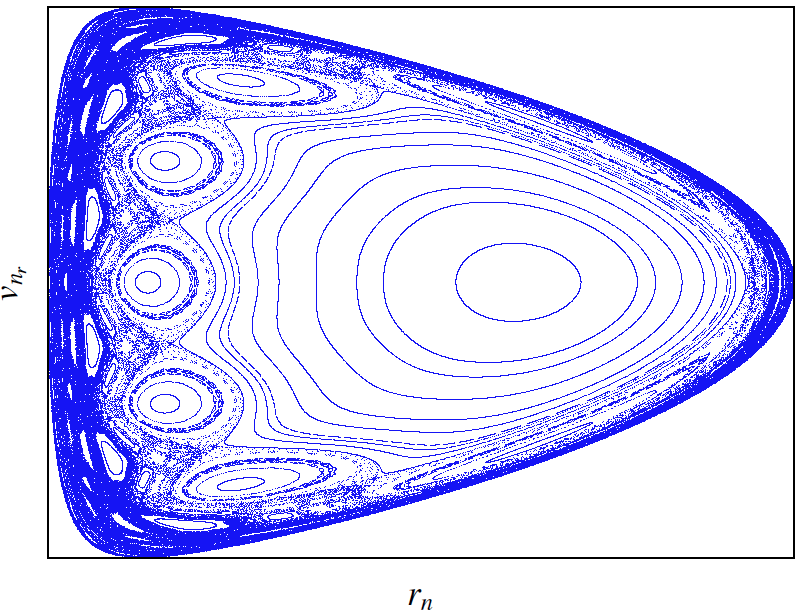}}\label{fig:6b}
\qquad
\subfloat[]{\includegraphics[height=4.5cm, width=4.4cm]{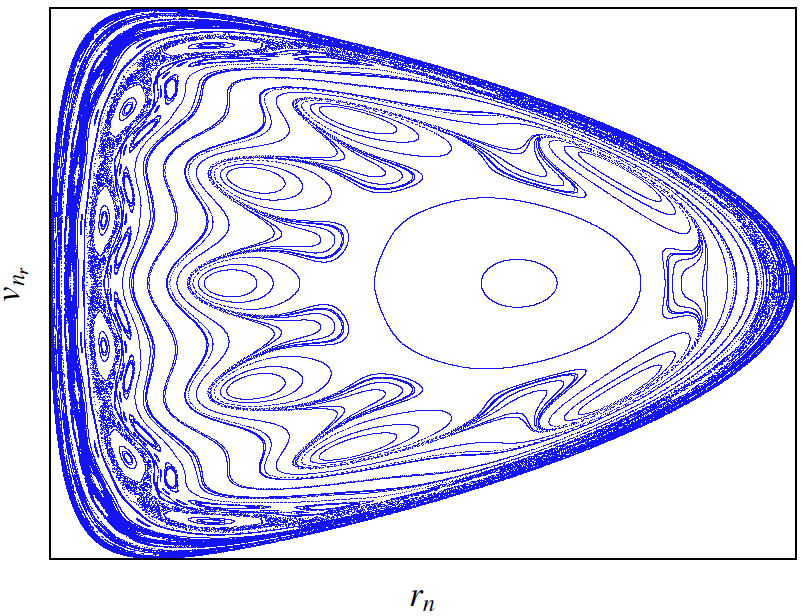}}\label{fig:6c}
\qquad
\subfloat[]{\includegraphics[height=4.5cm, width=4.4cm]{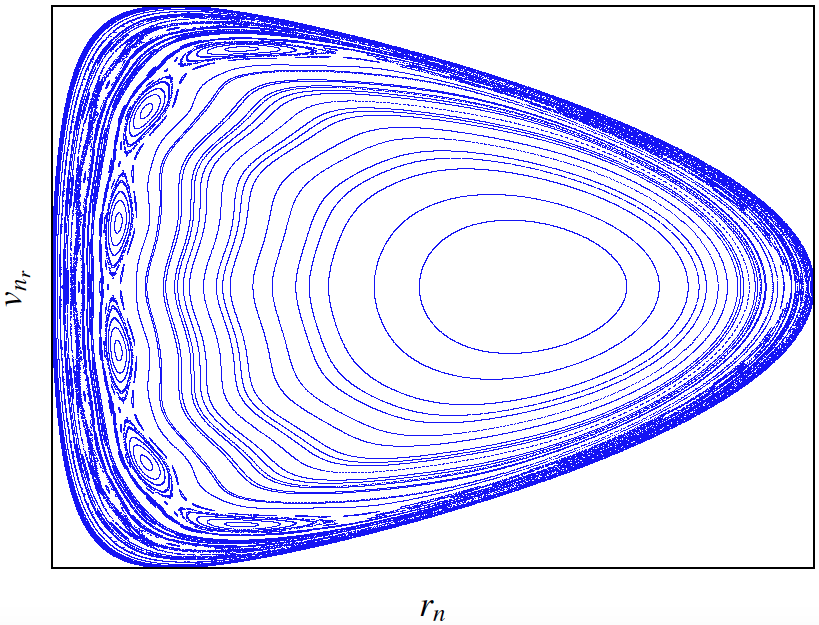}}\label{fig:6d}
\qquad
\subfloat[]{\includegraphics[height=4.5cm, width=4.4cm]{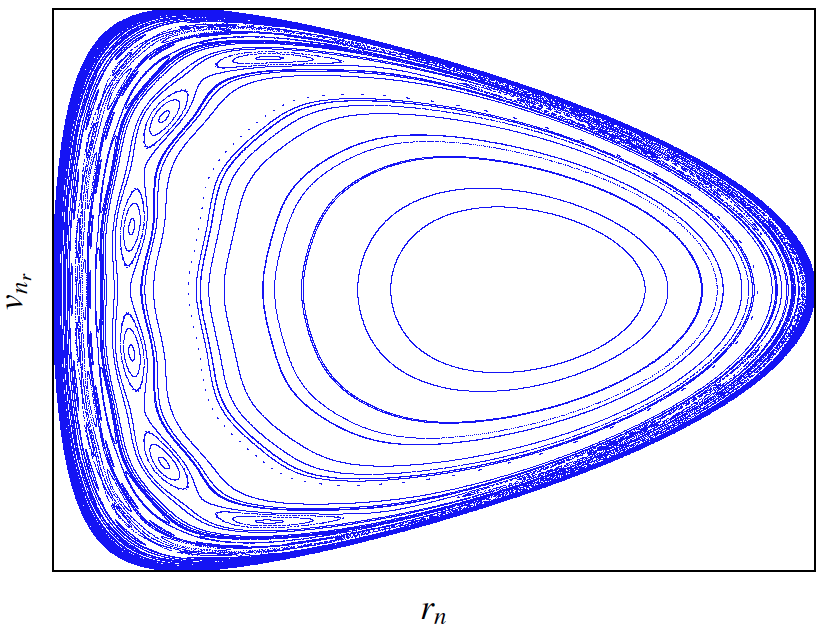}}\label{fig:6e}
\qquad 
\subfloat[]{\includegraphics[height=4.5cm, width=4.4cm]{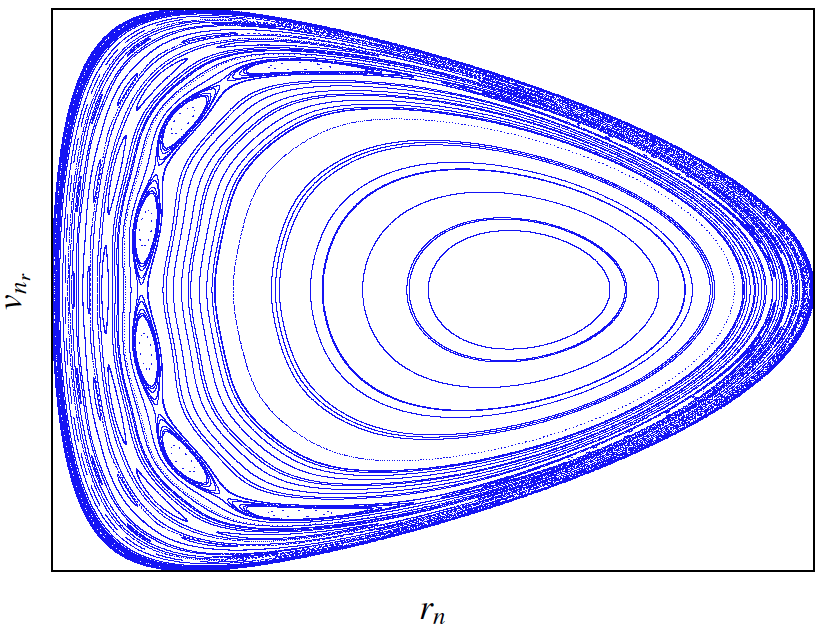}}\label{fig:6f}
\caption{SOS at $\theta=15^\circ$ for (a) $\ell=0.1$; (b) $\ell=0.13$; (c)  $\ell=0.16$; (d) $\ell=0.19$; (e) $\ell=0.22$; (f) $\ell=0.25$.}
\label{fig:transition}
\end{figure}

The transition between the small and large $\ell$-behavior is found to occur for the most part in the range $0.15\lesssim\ell\lesssim0.25$. In general the result of increasing $\ell$ is to reduce the amount of chaos in the phase space. In Fig.~\ref{fig:transition} we show this effect by plotting the Poincar\'{e} surface-of-section for different $\ell$ values at a fixed $\theta=15^\circ$. We see that the chaotic region present at $\ell=0.1$ has almost entirely disappeared by $\ell=0.25$, and has been replaced by stable island structures throughout the phase space. 

\section{Conclusions \& Outlook\label{sec:5}}
In this work we introduced a three-dimensional gravitational billiard consisting of a particle falling in a linear cone. We reduced this system to a two-dimensional area-preserving discrete map with two parameters. However, as demonstrated in section~\ref{sec:3.2} the $(r,v_r)$ mapping must be supplemented with the corresponding change in $\phi$ in order to fully understand the physical trajectory of the billiard. Thus, in this sense, the two-dimensional mapping does not always tell the whole story. We found several integrable limits of the system, computed the fixed point of the map and examined its linear stability as a function of parameter values. Due to the three-dimensional nature of the physical trajectory, this fixed point of the map does not in general describe a closed orbit in coordinate space. Next we investigated the phase space of the conic billiard in terms of the two parameters, $\theta$ and $\ell$. For small $\ell$ values we found that the system's phase space was qualitatively similar to the wedge, with a mixed phase space for $0^\circ<\theta<\theta_*$, which gives way to apparent ergodicity for $\theta>\theta_*$. In contrast to the wedge, however, this ergodicity does not hold for $\theta_*<\theta<90^\circ$; instead, stable periodic orbits re-appear, eventually leading to the integrable limit $\theta\rightarrow90^\circ$. As we increased $\ell$ we found the departure from the behavior of the wedge billiard to be dramatic. The relative amount of chaos in the phase space is reduced, with no simply connected region of chaos appearing for $\ell\gtrsim0.15$. The fixed point of the system plays a more significant role for large values of $\ell$, being stable for all cone angles with $\ell\gtrsim0.4$. This fixed point, together with surrounding KAM islands and additional periodic orbits, constitutes the majority of the phase space, with only a small chaotic band appearing for certain angles.\\\indent
In future work on the conic billiard we would like to include rotational effects, as well as examine driven versions of this system. One possible way to introduce time-dependence to the conic billiard would be to `spin' the cone sinusoidally. This would have the advantage that only the collision equations would need modification; computing the time of the next collision, which is in general a nontrivial numerical task, would be no more difficult than in the undriven conic system studied here. The conic billiard therefore offers an opportunity to examine the effects of rotation and time dependence on gravitational billiard systems. Another direction for future work is to consider different boundary surfaces. In two dimensions the parabolic billiard is the only known example of an integrable gravitational billiard~\cite{korsch1991}; in fact, it is possible to show that the corresponding three dimensional system, consisting of a gravitational billiard in a paraboloid, is also integrable. We plan to investigate this paraboloidal billiard in more detail in subsequent work. \\\indent
The conic billiard is distinct from the wedge primarily because of the dependence of the dynamics on the $z$-component of the particle's angular momentum. This is in contrast to most billiard systems, where the dynamics is governed \emph{only} by the shape of the boundary i.e., in this case the half-angle $\theta$ of the cone. The conic billiard is an example of a system where the initial condition of the particle (the $\phi$ component of the velocity) is folded into a parameter of the mapping characterizing the dynamics. Just as the properties of the wedge billiard have been experimentally confirmed using optical billiards~\cite{raizen1999}, the conic system studied here could likely be tested using ultracold atoms bouncing off laser beams.  

\appendix*
\section{Derivation of the velocity map}
Writing the collision equations \begin{equation} v_{{n+1}_r}^+=v_{{n+1}_r}^-,\qquad v_{{n+1}_\phi}^+=v_{{n+1}_\phi}^-,\qquad v_{{n+1}_\theta}^+=-v_{{n+1}_\theta}^-, \end{equation} in terms of the Cartesian velocity components results in a linear system of equations for the post-collision velocities $v_{{n+1}_x}^+,v_{{n+1}_y}^+$ and $v_{{n+1}_z}^+$ in terms of the corresponding pre-collision velocities $v_{{n+1}_x}^-,v_{{n+1}_y}^-$ and $v_{{n+1}_z}^-$. The solution of this system can be written \begin{align}\begin{split}\label{eq:velocity1} v_{{n+1}_x}^+ &=v_{{n+1}_x}^-(1-2\cos^2\phi_{n+1}\cos^2\theta)-v_{{n+1}_y}^-\sin2\phi_{n+1}\cos^2\theta+v_{{n+1}_z}^-\cos\phi_{n+1}\sin2\theta, \\  v_{{n+1}_y}^+ &=-v_{{n+1}_x}^-\sin2\phi_{n+1}\cos^2\theta+v_{{n+1}_y}^-(1-2\sin^2\phi_{n+1}\cos^2\theta)+v_{{n+1}_z}^-\sin\phi_{n+1}\sin2\theta,\\  v_{{n+1}_z}^+ &=v_{{n+1}_x}^-\cos\phi_{n+1}\sin2\theta+v_{{n+1}_y}^-\sin\phi_{n+1}\sin2\theta+v_{{n+1}_z}^-\cos2\theta.\end{split}\end{align} Since between collisions the billiard follows a simple parabolic trajectory, we have (recall that we set $g=\frac{1}{2}$ by a scaling transformation) \begin{equation}  v_{{n+1}_x}^-=v_{n_x},\qquad v_{{n+1}_y}^-=v_{n_y},\qquad v_{{n+1}_z}^-=-\frac{\tau_{n+1}}{2}+v_{n_z}.\end{equation} Using this in~\eqref{eq:velocity1} and writing the Cartesian components of the velocity in terms of the spherical components, the velocity map becomes (where we have abbreviated $v_{{n+1}_i}^+\equiv v_{{n+1}_i}$, and defined $\varphi_{n+1}\equiv \phi_{n+1}-\phi_n$) \begin{multline} v_{{n+1}_r} =v_{n_r}(\sin^2\theta\cos\varphi_{n+1}+\cos^2\theta)+v_{n_\theta}(\cos\varphi_{n+1}-1)\sin\theta\cos\theta+v_{n_\phi}\sin\varphi_{n+1}\sin\theta\\-\frac{1}{2}\tau_{n+1}\cos\theta, \end{multline} \begin{multline} v_{{n+1}_\theta} =v_{n_r}(1-\cos\varphi_{n+1})\sin\theta\cos\theta-v_{n_\theta}(\cos^2\theta\cos\varphi_{n+1}+\sin^2\theta)-v_{n_\phi}\sin\varphi_{n+1}\cos\theta\\-\frac{1}{2}\tau_{n+1}\sin\theta, \end{multline} \begin{equation} v_{{n+1}_\phi} =-v_{n_r}\sin\varphi_{n+1}\sin\theta-v_{n_\theta}\sin\varphi_{n+1}\cos\theta+v_{n_\phi}\cos\varphi_{n+1}.\end{equation} The difference in azimuthal angle $\varphi_{n+1}$ is easily found to be given by \[\varphi_{n+1}=\arctan\left[\frac{v_{n_\phi}\tau_{n+1}}{r_n\sin\theta+(v_{n_r}\sin\theta+v_{n_\theta}\cos\theta)\tau_{n+1}}\right],\] and after some algebraic manipulation this gives \begin{equation} \sin\varphi_{n+1}=\frac{v_{n_\phi}\tau_{n+1}}{\sqrt{[r_n\sin\theta+(v_{n_r}\sin\theta+v_{n_\theta}\cos\theta)\tau_{n+1}]^2+v_{n_\phi}^2\tau_{n+1}^2}}=\frac{v_{n_\phi}\tau_{n+1}}{r_{n+1}\sin\theta},\end{equation} and similarly \begin{equation} \cos\varphi_{n+1}=\frac{r_n+(v_{n_r}+v_{n_\theta}\cot\theta)\tau_{n+1}}{r_{n+1}}.\end{equation} In terms of the reduced variables \begin{equation} \rho\equiv r\sin\theta,\qquad u_r\equiv v_r\sin\theta,\qquad u_\theta\equiv v_\theta\cos\theta\end{equation} the $(\rho,u_r)$ mapping becomes \begin{equation} \rho_{n+1}^2 =\left[\left(u_{n_r}+u_{n_\theta}\right)^2+\frac{\ell^2}{\rho_n^2}\right]\tau_{n+1}^2+2\rho_n\left(u_{n_r}+u_{n_\theta}\right)+\rho_n^2, \end{equation} \begin{multline} u_{{n+1}_r}=\frac{\sin^2\theta}{\rho_{n+1}}\left\{\left[\left(u_{n_r}+u_{n_\theta}\right)^2+\frac{\ell^2}{\rho_n^2}\right]\tau_{n+1}+\rho_n\left(u_{n_r}+u_{n_\theta}\right)\right\}+u_{n_r}\cos^2\theta-u_{n_\theta}\sin^2\theta\\-\frac{1}{4}\tau_{n+1}\sin2\theta.\end{multline} The cubic equation for $\tau_{n+1}$ in terms of these reduced variables is \begin{multline} \tau_{n+1}^3+8(u_{n_\theta}\tan\theta-u_{n_r}\cot\theta)\tau_{n+1}^2\\+16\left[(\sec^2\theta-\csc^2\theta)u_{n_\theta}^2-\frac{\ell^2\cot^2\theta}{\rho_n^2}-2u_{n_r}u_{n_\theta}\csc^2\theta-\frac{\rho_n\cot\theta}{2}\right]\tau_{n+1} \\ -32\rho_nu_{n_\theta}\csc^2\theta=0.\end{multline}
\bibliography{Manuscript_1_bib}

\begin{thebibliography}{30}%
\makeatletter
\providecommand \@ifxundefined [1]{%
 \@ifx{#1\undefined}
}%
\providecommand \@ifnum [1]{%
 \ifnum #1\expandafter \@firstoftwo
 \else \expandafter \@secondoftwo
 \fi
}%
\providecommand \@ifx [1]{%
 \ifx #1\expandafter \@firstoftwo
 \else \expandafter \@secondoftwo
 \fi
}%
\providecommand \natexlab [1]{#1}%
\providecommand \enquote  [1]{``#1''}%
\providecommand \bibnamefont  [1]{#1}%
\providecommand \bibfnamefont [1]{#1}%
\providecommand \citenamefont [1]{#1}%
\providecommand \href@noop [0]{\@secondoftwo}%
\providecommand \href [0]{\begingroup \@sanitize@url \@href}%
\providecommand \@href[1]{\@@startlink{#1}\@@href}%
\providecommand \@@href[1]{\endgroup#1\@@endlink}%
\providecommand \@sanitize@url [0]{\catcode `\\12\catcode `\$12\catcode
  `\&12\catcode `\#12\catcode `\^12\catcode `\_12\catcode `\%12\relax}%
\providecommand \@@startlink[1]{}%
\providecommand \@@endlink[0]{}%
\providecommand \url  [0]{\begingroup\@sanitize@url \@url }%
\providecommand \@url [1]{\endgroup\@href {#1}{\urlprefix }}%
\providecommand \urlprefix  [0]{URL }%
\providecommand \Eprint [0]{\href }%
\providecommand \doibase [0]{http://dx.doi.org/}%
\providecommand \selectlanguage [0]{\@gobble}%
\providecommand \bibinfo  [0]{\@secondoftwo}%
\providecommand \bibfield  [0]{\@secondoftwo}%
\providecommand \translation [1]{[#1]}%
\providecommand \BibitemOpen [0]{}%
\providecommand \bibitemStop [0]{}%
\providecommand \bibitemNoStop [0]{.\EOS\space}%
\providecommand \EOS [0]{\spacefactor3000\relax}%
\providecommand \BibitemShut  [1]{\csname bibitem#1\endcsname}%
\let\auto@bib@innerbib\@empty
\bibitem [{\citenamefont {Birkhoff}(1927)}]{birkhoff1927}%
  \BibitemOpen
  \bibfield  {author} {\bibinfo {author} {\bibfnamefont {G.~D.}\ \bibnamefont
  {Birkhoff}},\ }\href@noop {} {\emph {\bibinfo {title} {Dynamical Systems}}}\
  (\bibinfo  {publisher} {American Mathematical Society},\ \bibinfo {address}
  {Providence, RI},\ \bibinfo {year} {1927})\BibitemShut {NoStop}%
\bibitem [{\citenamefont {Berry}(1981)}]{berry1981}%
  \BibitemOpen
  \bibfield  {author} {\bibinfo {author} {\bibfnamefont {M.~V.}\ \bibnamefont
  {Berry}},\ }\href {\doibase 10.1088/0143-0807/2/2/006} {\bibfield  {journal}
  {\bibinfo  {journal} {Eur. J. Phys.}\ }\textbf {\bibinfo {volume} {2}},\
  \bibinfo {pages} {91} (\bibinfo {year} {1981})}\BibitemShut {NoStop}%
\bibitem [{\citenamefont {Bunimovich}(1974)}]{bunimovich1974}%
  \BibitemOpen
  \bibfield  {author} {\bibinfo {author} {\bibfnamefont {L.~A.}\ \bibnamefont
  {Bunimovich}},\ }\href {\doibase 10.1007/BF01075700} {\bibfield  {journal}
  {\bibinfo  {journal} {Funct. Anal. Appl.}\ }\textbf {\bibinfo {volume} {8}},\
  \bibinfo {pages} {254} (\bibinfo {year} {1974})}\BibitemShut {NoStop}%
\bibitem [{\citenamefont {Bunimovich}(1979)}]{bunimovich1979}%
  \BibitemOpen
  \bibfield  {author} {\bibinfo {author} {\bibfnamefont {L.~A.}\ \bibnamefont
  {Bunimovich}},\ }\href {\doibase 10.1007/BF01197884} {\bibfield  {journal}
  {\bibinfo  {journal} {Commun. Math. Phys.}\ }\textbf {\bibinfo {volume}
  {65}},\ \bibinfo {pages} {295} (\bibinfo {year} {1979})}\BibitemShut
  {NoStop}%
\bibitem [{\citenamefont {Waalkens}\ \emph {et~al.}(1997)\citenamefont
  {Waalkens}, \citenamefont {Wiersig},\ and\ \citenamefont
  {Dullin}}]{waalkens1996}%
  \BibitemOpen
  \bibfield  {author} {\bibinfo {author} {\bibfnamefont {H.}~\bibnamefont
  {Waalkens}}, \bibinfo {author} {\bibfnamefont {J.}~\bibnamefont {Wiersig}}, \
  and\ \bibinfo {author} {\bibfnamefont {H.~R.}\ \bibnamefont {Dullin}},\
  }\href {\doibase 10.1006/aphy.1997.5715} {\bibfield  {journal} {\bibinfo
  {journal} {Ann. Phys}\ }\textbf {\bibinfo {volume} {260}},\ \bibinfo {pages}
  {50} (\bibinfo {year} {1997})}\BibitemShut {NoStop}%
\bibitem [{\citenamefont {Lehtihet}\ and\ \citenamefont
  {Miller}(1986)}]{miller86}%
  \BibitemOpen
  \bibfield  {author} {\bibinfo {author} {\bibfnamefont {H.~E.}\ \bibnamefont
  {Lehtihet}}\ and\ \bibinfo {author} {\bibfnamefont {B.~N.}\ \bibnamefont
  {Miller}},\ }\href {\doibase 10.1016/0167-2789(86)90080-1} {\bibfield
  {journal} {\bibinfo  {journal} {Physica D}\ }\textbf {\bibinfo {volume}
  {21}},\ \bibinfo {pages} {93} (\bibinfo {year} {1986})}\BibitemShut {NoStop}%
\bibitem [{\citenamefont {Richter}\ \emph {et~al.}(1990)\citenamefont
  {Richter}, \citenamefont {Scholz},\ and\ \citenamefont
  {Wittek}}]{richter1990}%
  \BibitemOpen
  \bibfield  {author} {\bibinfo {author} {\bibfnamefont {P.~H.}\ \bibnamefont
  {Richter}}, \bibinfo {author} {\bibfnamefont {H.~J.}\ \bibnamefont {Scholz}},
  \ and\ \bibinfo {author} {\bibfnamefont {A.}~\bibnamefont {Wittek}},\ }\href
  {\doibase 10.1088/0951-7715/3/1/004} {\bibfield  {journal} {\bibinfo
  {journal} {Nonlinearity}\ }\textbf {\bibinfo {volume} {3}},\ \bibinfo {pages}
  {45} (\bibinfo {year} {1990})}\BibitemShut {NoStop}%
\bibitem [{\citenamefont {Wojtkowski}(1990)}]{wojtkowski1990}%
  \BibitemOpen
  \bibfield  {author} {\bibinfo {author} {\bibfnamefont {M.~P.}\ \bibnamefont
  {Wojtkowski}},\ }\href {\doibase 10.1007/BF02125698} {\bibfield  {journal}
  {\bibinfo  {journal} {Comm. Math. Phys.}\ }\textbf {\bibinfo {volume}
  {126}},\ \bibinfo {pages} {507} (\bibinfo {year} {1990})}\BibitemShut
  {NoStop}%
\bibitem [{\citenamefont {Whelan}\ \emph {et~al.}(1990)\citenamefont {Whelan},
  \citenamefont {Goodings},\ and\ \citenamefont {Cannizzo}}]{whelan1990}%
  \BibitemOpen
  \bibfield  {author} {\bibinfo {author} {\bibfnamefont {N.~D.}\ \bibnamefont
  {Whelan}}, \bibinfo {author} {\bibfnamefont {D.~A.}\ \bibnamefont
  {Goodings}}, \ and\ \bibinfo {author} {\bibfnamefont {J.~K.}\ \bibnamefont
  {Cannizzo}},\ }\href {\doibase 10.1103/PhysRevA.42.742} {\bibfield  {journal}
  {\bibinfo  {journal} {Phys. Rev. A}\ }\textbf {\bibinfo {volume} {42}},\
  \bibinfo {pages} {742} (\bibinfo {year} {1990})}\BibitemShut {NoStop}%
\bibitem [{\citenamefont {Korsch}\ and\ \citenamefont
  {Lang}(1991)}]{korsch1991}%
  \BibitemOpen
  \bibfield  {author} {\bibinfo {author} {\bibfnamefont {H.~J.}\ \bibnamefont
  {Korsch}}\ and\ \bibinfo {author} {\bibfnamefont {J.}~\bibnamefont {Lang}},\
  }\href {\doibase 10.1088/0305-4470/24/1/015} {\bibfield  {journal} {\bibinfo
  {journal} {J. Phys. A: Math. Gen.}\ }\textbf {\bibinfo {volume} {24}},\
  \bibinfo {pages} {45} (\bibinfo {year} {1991})}\BibitemShut {NoStop}%
\bibitem [{\citenamefont {Szeredi}\ and\ \citenamefont
  {Goodings}(1993{\natexlab{a}})}]{1szeredi1993}%
  \BibitemOpen
  \bibfield  {author} {\bibinfo {author} {\bibfnamefont {T.}~\bibnamefont
  {Szeredi}}\ and\ \bibinfo {author} {\bibfnamefont {D.~A.}\ \bibnamefont
  {Goodings}},\ }\href {\doibase 10.1103/PhysRevE.48.3518} {\bibfield
  {journal} {\bibinfo  {journal} {Phys. Rev. E}\ }\textbf {\bibinfo {volume}
  {48}},\ \bibinfo {pages} {3518} (\bibinfo {year}
  {1993}{\natexlab{a}})}\BibitemShut {NoStop}%
\bibitem [{\citenamefont {Szeredi}\ and\ \citenamefont
  {Goodings}(1993{\natexlab{b}})}]{2szeredi1993}%
  \BibitemOpen
  \bibfield  {author} {\bibinfo {author} {\bibfnamefont {T.}~\bibnamefont
  {Szeredi}}\ and\ \bibinfo {author} {\bibfnamefont {D.~A.}\ \bibnamefont
  {Goodings}},\ }\href {\doibase 10.1103/PhysRevE.48.3529} {\bibfield
  {journal} {\bibinfo  {journal} {Phys. Rev. E}\ }\textbf {\bibinfo {volume}
  {48}},\ \bibinfo {pages} {3529} (\bibinfo {year}
  {1993}{\natexlab{b}})}\BibitemShut {NoStop}%
\bibitem [{\citenamefont {Ferguson}\ \emph {et~al.}(1999)\citenamefont
  {Ferguson}, \citenamefont {Miller},\ and\ \citenamefont
  {Thompson}}]{miller1999}%
  \BibitemOpen
  \bibfield  {author} {\bibinfo {author} {\bibfnamefont {M.~L.}\ \bibnamefont
  {Ferguson}}, \bibinfo {author} {\bibfnamefont {B.~N.}\ \bibnamefont
  {Miller}}, \ and\ \bibinfo {author} {\bibfnamefont {M.~A.}\ \bibnamefont
  {Thompson}},\ }\href {\doibase 10.1063/1.166467} {\bibfield  {journal}
  {\bibinfo  {journal} {Chaos}\ }\textbf {\bibinfo {volume} {9}},\ \bibinfo
  {pages} {841} (\bibinfo {year} {1999})}\BibitemShut {NoStop}%
\bibitem [{\citenamefont {Hartl}\ \emph {et~al.}(2013)\citenamefont {Hartl},
  \citenamefont {Miller},\ and\ \citenamefont {Mazzoleni}}]{miller2013}%
  \BibitemOpen
  \bibfield  {author} {\bibinfo {author} {\bibfnamefont {A.~E.}\ \bibnamefont
  {Hartl}}, \bibinfo {author} {\bibfnamefont {B.~N.}\ \bibnamefont {Miller}}, \
  and\ \bibinfo {author} {\bibfnamefont {A.~P.}\ \bibnamefont {Mazzoleni}},\
  }\href {\doibase 10.1103/PhysRevE.87.032901} {\bibfield  {journal} {\bibinfo
  {journal} {Phys. Rev. E}\ }\textbf {\bibinfo {volume} {87}},\ \bibinfo
  {pages} {032901} (\bibinfo {year} {2013})}\BibitemShut {NoStop}%
\bibitem [{\citenamefont {da~Costa}\ \emph {et~al.}(2015)\citenamefont
  {da~Costa}, \citenamefont {Dettmann},\ and\ \citenamefont
  {Leonel}}]{dacosta2015}%
  \BibitemOpen
  \bibfield  {author} {\bibinfo {author} {\bibfnamefont {D.~R.}\ \bibnamefont
  {da~Costa}}, \bibinfo {author} {\bibfnamefont {C.~P.}\ \bibnamefont
  {Dettmann}}, \ and\ \bibinfo {author} {\bibfnamefont {E.~D.}\ \bibnamefont
  {Leonel}},\ }\href {\doibase 10.1016/j.cnsns.2014.08.030} {\bibfield
  {journal} {\bibinfo  {journal} {Comm. in Non. Sci. and Num. Sim.}\ }\textbf
  {\bibinfo {volume} {22}},\ \bibinfo {pages} {731} (\bibinfo {year}
  {2015})}\BibitemShut {NoStop}%
\bibitem [{\citenamefont {Richter}\ \emph {et~al.}(1995)\citenamefont
  {Richter}, \citenamefont {Wittek}, \citenamefont {Kharlamov},\ and\
  \citenamefont {Kharlamov}}]{richter1995}%
  \BibitemOpen
  \bibfield  {author} {\bibinfo {author} {\bibfnamefont {P.~H.}\ \bibnamefont
  {Richter}}, \bibinfo {author} {\bibfnamefont {A.}~\bibnamefont {Wittek}},
  \bibinfo {author} {\bibfnamefont {M.~P.}\ \bibnamefont {Kharlamov}}, \ and\
  \bibinfo {author} {\bibfnamefont {A.~P.}\ \bibnamefont {Kharlamov}},\ }\href
  {\doibase 10.1515/zna-1995-0801} {\bibfield  {journal} {\bibinfo  {journal}
  {Z. Naturforschg.}\ }\textbf {\bibinfo {volume} {50a}},\ \bibinfo {pages}
  {693} (\bibinfo {year} {1995})}\BibitemShut {NoStop}%
\bibitem [{\citenamefont {Waalkens}\ \emph {et~al.}(1999)\citenamefont
  {Waalkens}, \citenamefont {Wiersig},\ and\ \citenamefont
  {Dullin}}]{waalkens1999}%
  \BibitemOpen
  \bibfield  {author} {\bibinfo {author} {\bibfnamefont {H.}~\bibnamefont
  {Waalkens}}, \bibinfo {author} {\bibfnamefont {J.}~\bibnamefont {Wiersig}}, \
  and\ \bibinfo {author} {\bibfnamefont {H.~R.}\ \bibnamefont {Dullin}},\
  }\href {\doibase 10.1006/aphy.1999.5936} {\bibfield  {journal} {\bibinfo
  {journal} {Ann. of Phys.}\ }\textbf {\bibinfo {volume} {276}},\ \bibinfo
  {pages} {64} (\bibinfo {year} {1999})}\BibitemShut {NoStop}%
\bibitem [{\citenamefont {Feldt}\ and\ \citenamefont
  {Olafsen}(2005)}]{feldt2005}%
  \BibitemOpen
  \bibfield  {author} {\bibinfo {author} {\bibfnamefont {S.}~\bibnamefont
  {Feldt}}\ and\ \bibinfo {author} {\bibfnamefont {J.~S.}\ \bibnamefont
  {Olafsen}},\ }\href {\doibase 10.1103/PhysRevLett.94.224102} {\bibfield
  {journal} {\bibinfo  {journal} {PRL}\ }\textbf {\bibinfo {volume} {94}},\
  \bibinfo {pages} {224102} (\bibinfo {year} {2005})}\BibitemShut {NoStop}%
\bibitem [{\citenamefont {Milner}\ \emph {et~al.}(2001)\citenamefont {Milner},
  \citenamefont {Hanssen}, \citenamefont {Campbell},\ and\ \citenamefont
  {Raizen}}]{raizen1999}%
  \BibitemOpen
  \bibfield  {author} {\bibinfo {author} {\bibfnamefont {V.}~\bibnamefont
  {Milner}}, \bibinfo {author} {\bibfnamefont {J.~L.}\ \bibnamefont {Hanssen}},
  \bibinfo {author} {\bibfnamefont {W.~C.}\ \bibnamefont {Campbell}}, \ and\
  \bibinfo {author} {\bibfnamefont {M.~G.}\ \bibnamefont {Raizen}},\ }\href
  {\doibase 10.1103/PhysRevLett.86.1514} {\bibfield  {journal} {\bibinfo
  {journal} {Phys. Rev. Lett}\ }\textbf {\bibinfo {volume} {86}},\ \bibinfo
  {pages} {1514} (\bibinfo {year} {2001})}\BibitemShut {NoStop}%
\bibitem [{\citenamefont {Holmes}(1982)}]{holmes1982}%
  \BibitemOpen
  \bibfield  {author} {\bibinfo {author} {\bibfnamefont {P.~J.}\ \bibnamefont
  {Holmes}},\ }\href@noop {} {\bibfield  {journal} {\bibinfo  {journal}
  {Journal of Sound and Vibration}\ }\textbf {\bibinfo {volume} {84}},\
  \bibinfo {pages} {173} (\bibinfo {year} {1982})}\BibitemShut {NoStop}%
\bibitem [{\citenamefont {Luck}\ and\ \citenamefont {Mehta}(1993)}]{luck1993}%
  \BibitemOpen
  \bibfield  {author} {\bibinfo {author} {\bibfnamefont {J.~M.}\ \bibnamefont
  {Luck}}\ and\ \bibinfo {author} {\bibfnamefont {A.}~\bibnamefont {Mehta}},\
  }\href@noop {} {\bibfield  {journal} {\bibinfo  {journal} {Phys. Rev. E}\
  }\textbf {\bibinfo {volume} {48}},\ \bibinfo {pages} {3988} (\bibinfo {year}
  {1993})}\BibitemShut {NoStop}%
\bibitem [{\citenamefont {Vogel}\ and\ \citenamefont {Linz}(2011)}]{vogel2011}%
  \BibitemOpen
  \bibfield  {author} {\bibinfo {author} {\bibfnamefont {S.}~\bibnamefont
  {Vogel}}\ and\ \bibinfo {author} {\bibfnamefont {S.~J.}\ \bibnamefont
  {Linz}},\ }\href {\doibase 10.1142/S0218127411028854} {\bibfield  {journal}
  {\bibinfo  {journal} {Int. J. of Bifurcation and Chaos}\ }\textbf {\bibinfo
  {volume} {21}},\ \bibinfo {pages} {869} (\bibinfo {year} {2011})}\BibitemShut
  {NoStop}%
\bibitem [{\citenamefont {Okninski}\ and\ \citenamefont
  {Radziszewski}(2009{\natexlab{a}})}]{okninski2009}%
  \BibitemOpen
  \bibfield  {author} {\bibinfo {author} {\bibfnamefont {A.}~\bibnamefont
  {Okninski}}\ and\ \bibinfo {author} {\bibfnamefont {B.}~\bibnamefont
  {Radziszewski}},\ }\href@noop {} {\bibfield  {journal} {\bibinfo  {journal}
  {Nonlinear Dynamics}\ }\textbf {\bibinfo {volume} {58}},\ \bibinfo {pages}
  {515} (\bibinfo {year} {2009}{\natexlab{a}})}\BibitemShut {NoStop}%
\bibitem [{\citenamefont {Okninski}\ and\ \citenamefont
  {Radziszewski}(2009{\natexlab{b}})}]{okninski2009-2}%
  \BibitemOpen
  \bibfield  {author} {\bibinfo {author} {\bibfnamefont {A.}~\bibnamefont
  {Okninski}}\ and\ \bibinfo {author} {\bibfnamefont {B.}~\bibnamefont
  {Radziszewski}},\ }in\ \href@noop {} {\emph {\bibinfo {booktitle} {Modeling,
  Simulation, and Control of Nonlinear Engineering Dynamical Systems}}}\
  (\bibinfo  {publisher} {Springer Berlin},\ \bibinfo {year} {2009})\ p.\
  \bibinfo {pages} {117}\BibitemShut {NoStop}%
\bibitem [{\citenamefont {Langer}\ and\ \citenamefont
  {Miller}(2015)}]{langer2015}%
  \BibitemOpen
  \bibfield  {author} {\bibinfo {author} {\bibfnamefont {C.~K.}\ \bibnamefont
  {Langer}}\ and\ \bibinfo {author} {\bibfnamefont {B.~N.}\ \bibnamefont
  {Miller}},\ }\href {\doibase 10.1063/1.4923747} {\bibfield  {journal}
  {\bibinfo  {journal} {Chaos}\ }\textbf {\bibinfo {volume} {25}},\ \bibinfo
  {pages} {073114} (\bibinfo {year} {2015})}\BibitemShut {NoStop}%
\bibitem [{\citenamefont {Fermi}(1949)}]{fermi1949}%
  \BibitemOpen
  \bibfield  {author} {\bibinfo {author} {\bibfnamefont {E.}~\bibnamefont
  {Fermi}},\ }\href {\doibase 10.1103/PhysRev.75.1169} {\bibfield  {journal}
  {\bibinfo  {journal} {Physical Review}\ }\textbf {\bibinfo {volume} {75}},\
  \bibinfo {pages} {1169} (\bibinfo {year} {1949})}\BibitemShut {NoStop}%
\bibitem [{\citenamefont {Ulam}(1961)}]{ulam1961}%
  \BibitemOpen
  \bibfield  {author} {\bibinfo {author} {\bibfnamefont {S.~M.}\ \bibnamefont
  {Ulam}},\ }in\ \href@noop {} {\emph {\bibinfo {booktitle} {Proc. Fourth
  Berkeley Symp. on Math. Statist. and Prob.}}},\ Vol.~\bibinfo {volume} {3}\
  (\bibinfo {year} {1961})\ pp.\ \bibinfo {pages} {315--320}\BibitemShut
  {NoStop}%
\bibitem [{\citenamefont {Gorski}\ and\ \citenamefont
  {Srokowski}(2006)}]{gorski2006}%
  \BibitemOpen
  \bibfield  {author} {\bibinfo {author} {\bibfnamefont {A.~Z.}\ \bibnamefont
  {Gorski}}\ and\ \bibinfo {author} {\bibfnamefont {T.}~\bibnamefont
  {Srokowski}},\ }\href@noop {} {\bibfield  {journal} {\bibinfo  {journal}
  {Acta. Phys. Pol. B}\ }\textbf {\bibinfo {volume} {37}},\ \bibinfo {pages}
  {2561} (\bibinfo {year} {2006})}\BibitemShut {NoStop}%
\bibitem [{Note1()}]{Note1}%
  \BibitemOpen
  \bibinfo {note} {We use $\theta $ as both the polar angle \protect \emph
  {and} the half-angle of the cone, because at a collision these angles must be
  equal; since we use a discrete map between collisions, the meaning of $\theta
  $ should be clear from context.}\BibitemShut {Stop}%
\bibitem [{\citenamefont {Greene}\ \emph {et~al.}(1981)\citenamefont {Greene},
  \citenamefont {Mckay}, \citenamefont {Vivaldi},\ and\ \citenamefont
  {Feigenbaum}}]{green1981}%
  \BibitemOpen
  \bibfield  {author} {\bibinfo {author} {\bibfnamefont {J.~M.}\ \bibnamefont
  {Greene}}, \bibinfo {author} {\bibfnamefont {R.~S.}\ \bibnamefont {Mckay}},
  \bibinfo {author} {\bibfnamefont {F.}~\bibnamefont {Vivaldi}}, \ and\
  \bibinfo {author} {\bibfnamefont {M.~J.}\ \bibnamefont {Feigenbaum}},\ }\href
  {\doibase 10.1016/0167-2789(81)90034-8} {\bibfield  {journal} {\bibinfo
  {journal} {Physica D}\ }\textbf {\bibinfo {volume} {3}},\ \bibinfo {pages}
  {468} (\bibinfo {year} {1981})}\BibitemShut {NoStop}%
\end{thebibliography}%

\end{document}